\documentclass[12pt]{article}
\pdfoutput=1
\usepackage{latexsym, amsmath, amssymb,  graphicx,  subfigure, simplewick}
\renewcommand{\thefootnote}{\fnsymbol{footnote}}

\usepackage{mathrsfs }
\usepackage{amsmath}
\usepackage{amsfonts}
\usepackage{eufrak}
\usepackage{eucal}
\numberwithin{equation}{section}
	
\def\doubleset#1#2{\bgroup%
\def\doit#1#2{%
\setbox\dblsetbox=\hbox{$\cstyle #1$}%
\raise#2\ht\dblsetbox\copy\dblsetbox%
\hskip-\wd\dblsetbox%
\raise-#2\ht\dblsetbox\box\dblsetbox}%
\mathchoice%
{\def\cstyle{\displaystyle}\doit#1#2}%
{\def\cstyle{\textstyle}\doit#1#2}%
{\def\cstyle{\scriptstyle}\doit#1#2}%
{\def\cstyle{\scriptscriptstyle}\doit#1#2}\egroup}
\def\underarrow#1{\vbox{\ialign{##\crcr$\hfil\displaystyle
 {#1}\hfil$\crcr\noalign{\kern1pt\nointerlineskip}$\longrightarrow$\crcr}}}

\def\IL{\relax{\rm I\kern-.18em L}}
\def\IH{\relax{\rm I\kern-.18em H}}
\def\IB{\relax{\rm I\kern-.18em B}}
\def\ID{\relax{\rm I\kern-.18em D}}
\def\IE{\relax{\rm I\kern-.18em E}}
\def\IF{\relax{\rm I\kern-.18em F}}

\def\IG{\relax\hbox{$\inbar\kern-.3em{\rm G}$}}
\def\IGa{\relax\hbox{${\rm I}\kern-.18em\Gamma$}}
\def\IH{\relax{\rm I\kern-.18em H}}
\def\II{\relax{\rm I\kern-.18em I}}
\def\IK{\relax{\rm I\kern-.18em K}}
\def\IP{\relax{\rm I\kern-.18em P}}
\def\IQ{\relax\hbox{$\inbar\kern-.3em{\rm Q}$}}

\def\hat{\widehat}
\def\CM {{\cal M}}

\def\inbar{\, \vrule height1.5ex width.4pt depth0pt}

\newbox\dblsetbox

\newlength{\extraspace}
\newlength{\extraspaces}
\newcommand{\be}{\begin{equation}
\addtolength{\abovedisplayskip}{\extraspaces}
\addtolength{\belowdisplayskip}{\extraspaces}
\addtolength{\abovedisplayshortskip}{\extraspace}
\addtolength{\belowdisplayshortskip}{\extraspace}}
\newcommand{\ee}{\end{equation}}

\newcommand{\ba}{\begin{eqnarray}
\addtolength{\abovedisplayskip}{\extraspaces}
\addtolength{\belowdisplayskip}{\extraspaces}
\addtolength{\abovedisplayshortskip}{\extraspace}
\addtolength{\belowdisplayshortskip}{\extraspace}}
\newcommand{\ea}{\end{eqnarray}}

\newcommand{\bd}{\begin{displaymath}
\addtolength{\abovedisplayskip}{\extraspaces}
\addtolength{\belowdisplayskip}{\extraspaces}
\addtolength{\abovedisplayshortskip}{\extraspace}
\addtolength{\belowdisplayshortskip}{\extraspace}}
\newcommand{\ed}{\end{displaymath}}

\newcounter{saveeqn}

\newcommand{\newsection}[1]{
\vspace{12mm} \pagebreak[3] \addtocounter{section}{1}
\setcounter{equation}{0} \setcounter{subsection}{0}
\noindent{\bf \thesection. #1} \nopagebreak
\medskip
\nopagebreak
\addcontentsline{toc}{section}{\thesection. #1}}

\newcommand{\newsubsection}[1]{
\vspace{0.8cm} \pagebreak[3] \addtocounter{subsection}{1}
\setcounter{subsubsection}{0}
\noindent{ \it \thesubsection. #1} \nopagebreak \vspace{2mm}
\nopagebreak
\addcontentsline{toc}{subsection}{\thesubsection. #1}}

\begin{document}
\addtolength{\baselineskip}{1.5mm}

\thispagestyle{empty}

\vbox{} \vspace{1.0cm}

\begin{center}
\centerline{\LARGE{\bf A Topological Chern-Simons Sigma Model and}}
\bigskip
\centerline{\LARGE{\bf New Invariants of Three-Manifolds}}

\vspace{1.5cm}

{\bf{Yuan Luo\footnote{ g0900733@nus.edu.sg} and Meng-Chwan~Tan\footnote{mctan@nus.edu.sg}}}
\\[0mm]
{\it Department of Physics,
National University of Singapore \\ 2 Science Drive 3, Singapore 117551}
\end{center}

\vspace{1.5 cm}

\centerline{\bf Abstract}\smallskip

We construct a topological Chern-Simons sigma model on a Riemannian three-manifold $M$ with gauge group $G$ whose hyperk\"ahler target space $X$ is equipped with a $G$-action. Via a perturbative computation of its partition function, we obtain \emph{new} topological invariants of $M$ that define \emph{new} weight systems which are characterized by both Lie algebra structure \emph{and} hyperk\"ahler geometry. In canonically quantizing the sigma model, we find that the partition function on certain $M$ can be expressed in terms of Chern-Simons knot invariants of $M$ and the intersection number of certain $G$-equivariant cycles in the moduli space of $G$-covariant maps from $M$ to $X$. We also construct supersymmetric Wilson loop operators, and via a perturbative computation of their expectation value, we obtain \emph{new} knot invariants of $M$ that define \emph{new} knot weight systems which are also characterized by both Lie algebra structure \emph{and} hyperk\"ahler geometry.

\newpage

\renewcommand{\thefootnote}{\arabic{footnote}}
\setcounter{footnote}{0}

\tableofcontents

\newpage
\newsection{Introduction, Summary and Acknowledgements}

The relevance of three-dimensional quantum field theory -- in particular, topological Chern-Simons gauge theory -- to the study of three-manifold invariants, was first elucidated in a seminal paper by Witten~\cite{Witten-Jones} in an attempt to furnish a three-dimensional interpretation of the Jones polynomial~\cite{Jones} of knots in three-space. Further developments~\cite{dror, Scott, E. Guad, Kont} along this direction culminated in the observation that certain three-manifold invariants can be expressed as weight systems whose weights depend on the Lie algebra structure which underlies the gauge group. Since these weights are naturally associated to Feynman diagrams via their relation to Chern-Simons theory, it meant that such three-manifold invariants have an alternative interpretation as Lie algebra-dependent graphical invariants. This provided a novel and interesting way to analyze them.    

It was then asked if there exist other three-manifold invariants that can be expressed as weight systems whose weights depend on something else other than Lie algebra structure. This question was answered positively by Rozansky and Witten several years later in~\cite{LR-Witten}, where they formulated a certain  three-dimensional supersymmetric topological sigma model with  a hyperk\"ahler target space -- better known today as the Rozansky-Witten sigma model -- and showed that one can, from its perturbative partition function, obtain such aforementioned three-manifold invariants whose weights depend not on Lie algebra structure but on hyperk\"ahler geometry. 
 
 Naturally, one may also ask if there exist even more exotic three-manifold invariants that can be expressed as weight systems whose weights depend on both Lie algebra structure \emph{and} hyperk\"ahler geometry. Clearly, the quantum field theory relevant to this question ought to be a hybrid of the Chern-Simons theory and the Rozansky-Witten sigma model -- a topological Chern-Simons sigma model if you will. Motivated by the formulation of such exotic three-manifold invariants among other things, the first example of a topological Chern-Simons sigma model -- also known as the Chern-Simons-Rozansky-Witten (CSRW) sigma model -- was constructed by Kapustin and Saulina in~\cite{Kap}. Shortly thereafter, a variety of other topological Chern-Simons sigma models was also constructed by Koh, Lee and Lee in~\cite{Koh}, following which, the CSRW model was reconstructed via the AKSZ formalism by K\"all\'en, Qiu and Zabzine in~\cite{zabzine}, where a closely-related (albeit non-Chern-Simons) BF-Rozansky-Witten sigma model was also presented. 
 
In these cited examples, the formulation and discussion of such exotic three-manifold invariants, if any at all, were rather abstract. Our main goal in this paper is to construct an appropriate Chern-Simons sigma model \footnote{ This model, just like the other CSRW-type models discussed in~\cite{Kap} and~\cite{Koh}, can be constructed by topologically twisting the theories discovered by Gaiotto and Witten in~\cite{G-W}. The theories constructed in~\cite{G-W} generalize $N=4\ d=3$ supersymmetric gauge theories which contain a Chern-Simons gauge field interacting with $N = 4$ hypermultiplets, by replacing the free hypermultiplets with a sigma model whose target space is a hyperK$\ddot{\text{a}}$hler manifold.} that would allow us to formulate and discuss, in a concrete and down-to-earth manner accessible to most physicists, such novel and exotic three-manifold invariants, their knot generalizations, and beyond. Let us now give a brief plan and summary of the paper.

\bigskip\noindent{\it A Brief Plan and Summary of the Paper}

In section 2, we construct from scratch, a topological Chern-Simons sigma model on a Riemannian three-manifold $M$ with gauge group $G$ whose hyperk\"ahler target space $X$ is equipped with a $G$-action, where $G$ is a compact Lie group with Lie algebra $\frak g$. Our model is a dynamically $G$-gauged version of the Rozansky-Witten sigma model, and it is closely-related to the Chern-Simons-Rozansky-Witten sigma model of Kapustin-Saulina: the Lagrangian of the models differ only by some mass terms for certain bosonic and fermionic fields. We also present a gauge-fixed version of the action, and discuss the (in)dependence of the partition function on the various coupling constants of the theory.  

In section 3, we compute perturbatively the partition function of the model. This is done by first expanding the quantum fields around points of stationary phase, and then evaluating the resulting Feynman diagram expansion of the path integral without operator insertions. Apart from obtaining \emph{new} three-manifold invariants which define \emph{new} weight systems whose weights are characterized by both the Lie algebra structure of $\frak g$  \emph{and} the hyperk\"ahler geometry of $X$, we also find that (i) the  one-loop contribution is a topological invariant of $M$ that ought to be related to a hybrid of the analytic Ray-Singer torsion of the flat and trivial connection on $M$, respectively; (ii) an ``equivariant linking number'' of knots in $M$ can be defined out of the propagators of certain fermionic fields.   

In section 4, we canonically quantize the time-invariant model in a neigborhood $\Sigma \times I$  of $M$, where $\Sigma$ is an arbitrary compact Riemann surface. We find that we effectively have a two-dimensional gauged sigma model on $\Sigma$, and that the relevant Hilbert space of states would be given by the tensor product of the Hilbert space of Chern-Simons theory on $M$ and the $G$-equivariant cohomology of the moduli space $\CM^\vartheta$ of $G$-covariant maps from $M$ to $X$. On three-manifolds $M^U$ which can be obtained from $M$ by a $U$-twisted surgery on $\Sigma = {\bf T}^2$, where $U$ is the mapping class group of $\Sigma$, the corresponding partition function $Z_X(M^U)$ can be expressed in terms of Chern-Simons knot invariants of $M$ and the intersection number of certain $G$-equivariant cycles in  $\CM^\vartheta$.

In section 5, we construct supersymmetric Wilson loop operators and compute perturbatively their expectation value. In doing so, we obtain \emph{new} knot invariants of $M$ that also define \emph{new} knot weight systems whose weights are characterized by both the Lie algebra structure of $\frak g$ \emph{and} the hyperk\"ahler geometry of $X$.

\bigskip\noindent{\it Acknowledgements}

We would like to thank D.~Bar-Natan, M.~Brion, A. Kapustin, L. Rozansky,  E. Witten and J.~Yagi, for useful exchanges. 

This work is supported in part by the NUS Startup Grant. 

\newsection{A Topological Chern-Simons Sigma Model}

\newsubsection{The Fields and the Action}

We would like to construct a topological Chern-Simons (CS) sigma model that is a dynamically $G$-gauged version of the Rozansky-Witten (RW) sigma model on $M$ with target space $X$, where $M$ is a three-dimensional Riemannian manifold with local coordinates $x^{\mu}$, $\mu=1, 2, 3$, and  $X$ is a hyperk\"ahler manifold of complex dimension ${\rm dim}_{\mathbb{C}}X=2n$ which admits an action of a compact Lie group $G$. Let $\{V_{a}\}$ where $a=1, 2, \cdots, {\rm dim} \, G$, be the set of Killing vector fields on $X$ which correspond to this $G$-action; they can be viewed as sections of $TX\otimes \mathfrak{g}^{*}$,  where $TX$ is the tangent bundle of $X$, while $\mathfrak{g}$ is the Lie algebra of $G$. If we denote the local complex coordinates of $X$ as  $(\phi^{I}, \phi^{\bar{I}})$, where $I, \bar{I}=1, \cdots, 2n$, one can also write these vector fields as
\begin{equation}
V_{a}=V_{a}^{I}\partial_{I}+V_{a}^{\bar{I}}\partial_{\bar{I}}.  \notag
\end{equation}
Note that the $V_{a}$'s satisfy the Lie algebra
\begin{equation}
[V_{a}, V_{b}]=f^{c}_{ab}V_{c},           \notag
\end{equation}
where the $f^{c}_{ab}$'s are the structure constants of $\mathfrak{g}$. Therefore,  $\phi^{I}$ and $\phi^{\bar{I}}$ must transform under the $G$-action as
\begin{equation}
\delta_{\epsilon}\phi^{I}=\epsilon^{a}V_{a}^{I},  \quad  \delta_{\epsilon}\phi^{\bar{I}}=\epsilon^{a}V_{a}^{\bar{I}}.    \notag
\end{equation}

In order for $G$ to be a global symmetry of $X$, it is necessary and sufficient that (i) for all $a$, the $V_{a}$'s are holomorphic or anti-holomorphic; (ii) the symplectic structure of $X$  is preserved by the $G$-action associated with the $V_{a}$'s. If the k\"ahler form on $X$ is also preserved by the $G$-action, locally, there would exist moment maps $\mu_{+}, \mu_{-}, \mu_{3}:X \to \mathfrak{g}^{*}$, where
\begin{equation}
d\mu_{+a}=-i_{V_{a}}(\Omega),\quad d\mu_{-a}=-i_{V_{a}}(\bar{\Omega}),\quad d\mu_{3a}=-i_{V_{a}}(J).
\end{equation}
Here, $\Omega=\frac{1}{2}\Omega_{IJ}d\phi^{I}\wedge d\phi^{J}$ is the holomorphic symplectic form on $X$;  $J=ig_{I\bar{K}}d\phi^{I}\wedge d\phi^{\bar{K}}$ is the k\"ahler form on $X$; $g_{I\bar{K}}$ is the metric on $X$; and $i_{V}(\omega)$ stands for the inner product of the vector field $V$ with the differential form $\omega$. The moment maps $\mu_{+}, \mu_{-}, \mu_{3}$ are assumed to exist globally (which is automatically the case if $X$ is simply-connected), and $\mu_{+}$ is holomorphic  while $\mu_{-}=\bar{\mu}_{+}$ is antiholomorphic.  $\mu_{+}$ also satisfies
\begin{equation}
\label{PB-mu}
\{\mu_{+a}, \mu_{+b}\}=-f^{c}_{ab}\mu_{+c},
\end{equation}
where the curly brackets are the Poisson brackets with respect to $\Omega_{IJ}$. Similar formulas hold for $\mu_{-}$ and $\mu_{3}$. We further assume that $X$ is such that
\begin{equation}
\mu_{+}\cdot \mu_{+}=\kappa^{ab}\mu_{+a}\mu_{+b}=0,
\label{moment m}
\end{equation}
because this condition is necessary for the supersymmetry transformation defined later to be nilpotent on gauge-invariant obeservables. Note that in (\ref{moment m}), $\kappa^{ab}$ is the inverse of the $G$-invariant nondegenerate symmetric bilinear  form $\kappa_{ab}$ on $\mathfrak{g}$, where
\begin{equation}
\kappa_{ad}f^{d}_{bc}+\kappa_{bd}f^{d}_{ac}=0.
\end{equation}

Now, the fields of a $G$-gauged version of the RW sigma model ought to be given by
\begin{equation}
\textrm{bosonic}: \phi^{I},  \phi^{\bar{I}},   A^{a}_{\mu}; \quad  \textrm{fermionic}: \eta^{\bar{I}},  \chi^{I}_{\mu},
\end{equation}
where $I,  \bar{I}=1, \cdots, 2n$; $\mu=1, 2, 3$; and $a=1, \cdots {\rm dim} \, G$. The gauge field $A$ is a connection one-form on a principal $G$-bundle $\varepsilon$ over $M$. With respect to an infinitesimal gauge transformation with parameter $\epsilon^{a}(x)$, it should transform as
\begin{equation}
\delta_{\epsilon}A^{a}=-(d\epsilon^{a}-f^{a}_{bc}A^{b}\epsilon^{c})=-D\epsilon^{a}.
\end{equation}

Since $G$ acts on $X$, the bosonic fields $\phi^{I}, \phi^{\bar{I}}$ must be sections of a fiber bundle over $M$ associated with $\varepsilon$, whose typical fiber is $X$. Denote this bundle as $X_{\varepsilon}$. Then, the connection $A$ also defines a nonlinear connection on $X_{\varepsilon}$ where locally, it can be thought of as a one-form on $M$ with values in the Lie algebra of vector fields on $X$, i.e., $A = A^{a}V_{a}$. This means that we can write the covariant differentials  of $\phi^{I}$ and $\phi^{\bar{I}}$ as
\begin{equation}
D\phi^{I}=d\phi^{I}+A^{a}V^{I}_{a},\quad D\phi^{\bar{I}}=d\phi^{\bar{I}}+A^{a}V^{\bar{I}}_{a}.         \notag
\end{equation}

As for the fermionic fields, $\chi^{I}_{\mu}$ are components of a one-form $\chi^{I}$ on $M$ with values in the pullback $\phi^{*}(T_{X_{\varepsilon}})$, where $T_{X_{\varepsilon}}$ is the $(1,0)$ part of the fiberwise-tangent bundle of $X_{\varepsilon}$, while $\eta^{\bar{I}}$ is a zero-form on $M$ with values in the pullback $\phi^{*}(\bar{T}_{X_{\varepsilon}})$ of the complex-conjugate bundle $\bar{T}_{X_{\varepsilon}}$.

From the above expressions, it is clear that the data of the Lie group $G$ and the hyperk\"ahler geometry of $X$ are inextricably connected.  This connection will allow us to obtain \emph{new} three-manifold invariants which depend on both $G$ and $X$, as we will show in the next section.

\bigskip\noindent{\it The Action}

At any rate, let us now construct the action of the model. Let us assign to the fields $\phi$, $\chi$, $\eta$ and $A$, the $U(1)$ $R$-charge $0, -1, 1$ and $0$, respectively. Let us also define the following supersymmetry transformation of the fields under a scalar supercharge $Q$:
\begin{eqnarray}
\notag
&&\delta_{Q}A_{a}=\chi^{K}\partial_{K}\mu_{+a},
\\ \notag
&&\delta_{Q}\phi^{I}=0,
\\
&&\delta_{Q}\phi^{\bar{I}}=\eta^{\bar{I}},
\\ \notag
&&\delta_{Q}\chi^{I}=D\phi^{I},
\\ \notag
&&\delta_{Q}\eta^{\bar{I}}=-\bar{\xi}^{\bar{I}},
\label{BRST}
\end{eqnarray}
where
\begin{equation}
\xi^{I}=V^{I}\cdot \mu_{-}, \quad \bar{\xi}^{\bar{I}}=V^{\bar{I}}\cdot \mu_{+}.
\end{equation}
Here, the scalar supercharge $Q$ is defined to have $R$-charge $+1$, while the moment maps $\mu_{\pm}$ are defined to have $R$-charge $\pm 2$. Notice then that spin and $R$-charge are conserved in the above relations, as required.

From (\ref{BRST}), we find that $\delta^{2}_{Q}$ is a gauge transformation with parameter $\epsilon^{a}=-\kappa^{ab}\mu_{+b}$:
\begin{eqnarray}
\notag
&&\delta^{2}_{Q}A^{a}=\kappa^{ab}(d\mu_{+b}+f^{d}_{cb}A^{c}\mu_{+d}),  \\
&&\delta^{2}_{Q}\phi^{I}=0, \quad \delta^{2}_{Q}\phi^{\bar{I}}=-V^{\bar{I}}\cdot \mu_{+}, \\
&&\delta^{2}_{Q}\chi^{I}=-\chi^{J}\partial_{J}V^{I}\cdot \mu_{+}, \quad \delta^{2}_{Q}\eta^{\bar{I}}=-\eta^{\bar{J}}\partial_{\bar{J}}V^{\bar{I}}\cdot \mu_{+}.       \notag
\label{Q2 gauge}
\end{eqnarray}
Note that to compute this, we have used $V^{K}_{a}\Omega_{KJ}V^{J}_{b}=f^{c}_{ab}\mu_{+c}$ and $V^{I}\cdot \mu_{+}=0$.

Thus, an example of a Q-invariant action $S$ would be
\begin{eqnarray}
S &&= \int_{M} (L_{cs}+L_{1}+L_{2}),\\ \notag
L_{cs}&&= {\rm Tr}(A\wedge dA+\frac{2}{3}A\wedge A\wedge A),\\ \notag
L_{1} &&= \delta_{Q}(g_{I\bar{K}}\chi^{I}\wedge \ast D\phi^{\bar{K}})  \\ \notag
        &&=g_{I\bar{K}}(D\phi^{I}\wedge \ast D\phi^{\bar{K}}-\chi^{I}\wedge \ast D\eta^{\bar{K}}),\\ \notag
L_{2}&&=\frac{1}{2}\Omega_{IJ}(\chi^{I}\wedge D\chi^{J}+\frac{1}{3}R^{J}_{KL\bar{M}}\chi^{I}\wedge \chi^{K}\wedge \chi^{L}\wedge \eta^{\bar{M}}),\notag
\end{eqnarray}
where $\ast$ denotes the Hodge star operator on differential forms on $M$ with respect to its Riemannian metric $h_{\mu\nu}$; `Tr' denotes a suitably-normalized invariant quadratic form on  $\frak g$; the covariant derivatives are given by
\begin{equation}
D\phi^{I}=d\phi^{I}+A\cdot V^{I},\quad D\chi^{I}=\nabla\chi^{I}+A\cdot\nabla_{K}V^{I}\chi^{K},\quad
D\eta^{\bar{I}}=\nabla\eta^{\bar{I}}+A\cdot\nabla_{\bar{K}}V^{\bar{I}}\eta^{\bar{K}};   \notag
\end{equation}
$\nabla$ involves the Levi-Civita connection on $X$, where
\begin{eqnarray}
\notag
\nabla\chi^{I}=d\chi^{I}+\Gamma^{I}_{JK}d\phi^{J}\wedge\chi^{K},\quad
\nabla\eta^{\bar{I}}=d\eta^{\bar{I}}+\Gamma^{\bar{I}}_{\bar{J}\bar{K}}d\phi^{\bar{J}}\wedge\eta^{\bar{K}},
\\
 \nabla_{K} V^{I}=\partial_{K}V^{I}+\Gamma^{I}_{KJ}V^{J},\quad
  \nabla_{\bar{K}} V^{\bar{I}}=\partial_{\bar{K}}V^{\bar{I}}+\Gamma^{\bar{I}}_{\bar{K}\bar{J}}V^{\bar{J}}; \notag
\end{eqnarray}
and $R^{J}_{KL\bar{M}}$ denotes the curvature tensor of the Levi-Civita connection on $X$, where
\begin{equation}
R^{J}_{KL\bar{M}}=\frac{\partial \Gamma^{J}_{KL}}{\partial\phi^{\bar{M}}},\quad  \Gamma^{I}_{JK}=(\partial_{J}g_{K\bar{M}})g^{I\bar{M}} . \notag
\end{equation}

\newsubsection{Gauge-Fixing}

One of our main objectives in this paper is to compute the partition function of the model. To do so, we need to gauge-fix the model. This can be done as follows.

Define the total BRST transformation
\begin{equation}\delta_{\hat{Q}}=\delta_{Q}+\delta_{FP}, \notag \end{equation}
where $\delta_{FP}$ is the usual Faddeev-Popov BRST operator with $R$-charge $+1$. The total BRST transformation $\delta_{\hat{Q}}$ must be nilpotent, while $\delta_{Q}$ is nilpotent only up to a gauge transformation.

We then extend the theory by introducing fermionic Faddev-Popov ghost and anti-ghost fields $c^{a}$, $\bar{c}_{a}$, as well as bosonic Lagrangian multiplier fields $B_{a}$. $c,\ \bar{c},\ B$ are defined to have $R$-charge $1$, $-1$ and $0$, respectively. $c$ takes values in $\mathfrak{g}$, while $\bar{c}$ and $B$ take values in the dual Lie algebra $\mathfrak{g}^{*}$. By conservation of spin and $R$-charge, the total BRST operator $\hat{Q}$ should act on the fields as
\begin{eqnarray}
\label{BRSTgf}
&&\delta_{\hat{Q}}A_{a}=dc_{a}-f_{abd}A^{b}c^{d}+\chi^{K}\partial_{K}\mu_{+a},\notag
\\
&&\delta_{\hat{Q}}\phi^{I}=-V^{I} \cdot c, \notag
\\
&&\delta_{\hat{Q}}\phi^{\bar{I}}=\eta^{\bar{I}}-V^{\bar{I}} \cdot c, \notag
\\
&&\delta_{\hat{Q}}\chi^{I}=D\phi^{I}+(\partial_{J}V^{Ia})\chi^{J}c_{a},
\\
&&\delta_{\hat{Q}}\eta^{\bar{I}}=\-\bar{\xi}^{\bar{I}}+(\partial_{\bar{J}}V^{\bar{I}a})\eta^{\bar{J}}c_{a}, \notag
\\
&&\delta_{\hat{Q}}c^{a}=-\kappa^{ab}\mu_{+b}+\frac{1}{2}f^{a}_{bc}c^{b}c^{c}, \notag
\\
&&\delta_{\hat{Q}}\bar{c}\ =B, \notag
\\
&&\delta_{\hat{Q}}B=0. \notag
\end{eqnarray}

It's easy to show that $\delta^2_{\hat{Q}}=0$ on the fields. The ${\hat{Q}}$-invariant gauge-fixed action $S$ would then be
\begin{eqnarray}
\notag
&&S= \int_{M} (L_{cs}+L_{1}+L_{2}),\\ \notag
&&L_{cs}= {\rm Tr} (A\wedge dA+\frac{2}{3}A\wedge A\wedge A),\\
&&L_{1}= \delta_{\hat{Q}}(g_{I\bar{K}}\chi^{I}\wedge \ast D\phi^{\bar{K}}+\bar{c}_{a}f^{a}) \\ \notag
&&\ \ \ \  =g_{I\bar{K}}(D\phi^{I}\wedge \ast D\phi^{\bar{K}}-\chi^{I}\wedge \ast D\eta^{\bar{K}})
        +B_{a}f^{a}-\bar{c}_{a}\delta_{\hat{Q}}f^{a}, \\ \notag
&&L_{2} =\frac{1}{2}\Omega_{IJ}(\chi^{I}\wedge D\chi^{J}+\frac{1}{3}R^{J}_{KL\bar{M}}\chi^{I}\wedge \chi^{K}\wedge \chi^{L}\wedge \eta^{\bar{M}}),\notag
\end{eqnarray}
where $f^a$ is some $\frak g$-valued function; $\int_{M} L_{cs}$ and $\int_{M} L_{2}$ are manifestly independent of the metric of $M$; while $L_{1} = \{\hat{Q}, \dots \}$ is an exact form of the total BRST operator $\hat Q$. Since the metric dependence of the action is of the form $\{\hat{Q}, \dots \}$, the partition function, and also the correlation functions of $\hat{Q}$-closed operators, are metric independent. In this sense, the theory is topologically invariant.

Notice that the transformation on the ghost field $c$ is not standard. The standard ghost field transformation just involves the usual $\delta_ {FP}$ variation, while $c$ also gets transformed by $\delta_{Q}$:
\begin{equation}
\delta_{Q}c^{a}=\kappa^{ab}\mu_{+b}.
\end{equation}
This fact makes the part of the action involving ghost and anti-ghost fields non-standard. For example, if we choose the Lorentz gauge $f^{a}=\partial^{\mu}A_{\mu}^{a}$, the action contains the term $\bar{c}^{a}\partial^{\mu}(\chi^{K}\mu_{+a})$ where the anti-ghost field $\bar{c}^{a}$ is coupled to the `matter' fermion $\chi^{K}$.

\newsubsection{About the Coupling Constants}

Before we end this section,  let us discuss the coupling constants of the theory as it would prove useful to do so when we carry out our computation of the partition function and beyond in the rest of the paper. 

To this end, note that the partition function can be written as
\begin{equation}
Z=\int{D\phi D\eta D\chi DA Dc D\bar{c} DB \ \exp \left(-\int_{M}(k_{cs}L_{cs}+k_{1}L_{1}+k_{2}L_{2})\sqrt{h}d^{3}x \right)},
\end{equation}
where $k_1$, $k_2$ and $k_{cs}$ are the possible coupling constants of the theory. As
\begin{equation}
\frac{\delta Z}{\delta k_{1}}=\langle \delta_{\hat{Q}}\mathcal{O} \rangle=0,
\end{equation}
the partition function should not depend on $k_{1}$.

Let us now rescale the fields as follows:
\begin{equation}
\eta\rightarrow\lambda\eta,\quad \chi\rightarrow\lambda^{-1}\chi,\quad \bar{c}\rightarrow\lambda\bar{c},\quad c\rightarrow\lambda^{-1}c,
\end{equation}
whence
\begin{equation}
k_{1}L_{1}\rightarrow k_{1}L_{1},\quad  k_{2}L_{2}\rightarrow\lambda^{-2}k_{2}L_{2}.
\end{equation}
As the field rescaling should not change the theory, the partition function should not depend on $k_{2}$ either. Thus, let us just write
\begin{equation}
k_{1}=k_{2}=k.
\end{equation}

That being said, our partition function $\it{does}$ depend on the coupling constant $k_{cs}$. Moreover, because of the requirement of gauge invariance~\cite{Witten-Jones}, $k_{cs}$ ought to be quantized as 
\begin{equation}
k_{cs}=\frac{m}{2\pi}; \quad m=1, 2, 3 \dots
\end{equation}

Hence, we have \emph{two} physically distinct coupling constants in our theory. This should come as no surprise since our theory is actually a combination of a Schwarz- and Witten-type topological field theory.

\newsection{The Perturbative Partition Function and New Three-Manifold Invariants}

\newsubsection{The Perturbative Partition Function}

Let us now proceed to discuss the partition function of the gauged sigma model in the perturbative limit. To this end, recall from the last section that the partition function depends on the coupling $k_{cs}$. Hence, the perturbative limit of the (CS part of the) model is the same as its large $k_{cs}$ limit. Moreover, because the partition function is independent of $k$, we can choose $k_{1}=k_{2}=k$ as large as we want. Altogether, this means that the perturbative partition function would be given by a sum of contributions centered around the points of stationary phase characterized by
\begin{equation}
\frac{\delta{L_{cs}}}{\delta{A}}=dA+[A, A]=0,
\end{equation}
which are the \emph{flat connections}, and
\begin{equation}
\frac{\delta S}{\delta \phi}=0\to D^{\mu}\phi=0,
\end{equation}
which are the \emph{covariantly constant} maps from $M$ to $X$.

 Thus, where the perturbative partition function is concerned, we can expand the gauge field $A$ around the flat connection $A^\vartheta_0$ as
\begin{equation}
\label{A-expand}
A^{a}_{\mu}(x)=A_{0\mu}^{\vartheta a}(x)+\tilde{A}^{a}_{\mu}(x),
\end{equation}
and the bosonic scalar fields $\phi$ around the covariantly constant map $\phi_0$ as
\begin{equation}
\label{phi-expand}
\phi^{I}(x)=\phi^{I}_{0}(x)+\varphi^{I}(x),\quad \phi^{\bar{I}}(x)=\phi^{\bar{I}}_{0}(x)+\varphi^{\bar{I}}(x),
\end{equation}
where
\begin{equation}
\label{cov phi}
D^{\mu}\phi_{0}^{I}=\partial^{\mu}\phi_{0}^{I}+A_{0}^{\vartheta a \mu }V^{I}_a(\phi_{0})=0, \qquad D^{\mu}\phi_{0}^{\bar I}=\partial^{\mu}\phi_{0}^{\bar I}+A_{0}^{\vartheta a \mu }V^{\bar I}_a(\phi_{0})=0. 
\end{equation}

Note that (\ref{A-expand}) means that we can write
\begin{equation}
L_{cs}=L_{cs}(A^\vartheta_{0})+ \tilde{A}\wedge d\tilde{A}+\tilde{A}\wedge[A^\vartheta_{0}, \tilde{A}]+\frac{2}{3}\tilde{A}\wedge \tilde{A}\wedge \tilde{A}.
\end{equation}
Let $\mathcal{M^\vartheta}$ be the space of physically distinct $\phi_0$'s which satisfy (\ref{cov phi}) for some flat connection   $A_{0}^{\vartheta}$. Assuming that the flat connection $A_{0}^{\vartheta}$ is isolated,\footnote{This would indeed be the case if $H^1(M, E) = 0$, where $E$ is a flat bundle determined by $A_{0}$.\label{H1M}} we can then write our perturbative partition function as
\begin{equation}
Z = k^{2n}\sum_{A^{\vartheta}_{0}}
\left(
e^{-\int_{M}L_{cs}(A^{\vartheta}_{0})}
 \int_{\mathcal{M^\vartheta}} \prod_{I=1}^{2n} d\phi^{I}_{0}
 \prod_{\bar I=1}^{2n} d\phi^{\bar{I}}_{0} 
\int D\varphi D\chi D\eta D\tilde{A} Dc D\bar{c} DB \, e^{-S_{A^\vartheta_{0},\phi_{0}}}
\right).
\end{equation}
Here, $k^{2n}$ is the normalization factor carried by the $2n$ bosonic zero modes $\phi_0$, and $\int_{M}L_{cs}(A^{\vartheta}_{0}) + S_{A^\vartheta_{0},\phi_{0}}$ is the total action expanded around $A^\vartheta_{0}$ and $\phi_{0}$.

In the total action expanded around the flat gauge field $A^\vartheta_{0}$ and the covariantly constant bosonic scalar fields $\phi^{I, \bar I}_{0}$, we have
\begin{eqnarray}
\label{Dphi}
D_\mu\phi^{I} &&=\partial_\mu\phi_{0}^{I}+ \partial_\mu \varphi^{I}+(A^{\vartheta a}_{0 \mu}+\tilde{A}^{a}_\mu) \{V^I_a(\phi_{0})+\varphi^{J}\partial_{J}V^{I}_a(\phi_{0})+\frac{\varphi^{J}\varphi^{K}}{2}\partial_{J}\partial_{K}V^{I}_a(\phi_{0})+\cdots \}
 \nonumber \\
&&=\partial_\mu \varphi^{I} +A^{\vartheta a}_{0 \mu}\varphi^{J}\partial_{J}V^{I}_a +\tilde{A}^a_\mu V^{I}_a +\tilde{A}^a_\mu \varphi^{J}\partial_{J}V^{I}_a +(A^{\vartheta a}_{0 \mu}+\tilde{A}^a_\mu)(\frac{\varphi^{J}\varphi^{K}}{2}\partial_{J}\partial_{K}V^{I}_a)+\cdots, \nonumber \\
\end{eqnarray}
since $D_\mu \phi^I_0 = \partial_\mu \phi^I_{0} + A^{\vartheta a}_{0 \mu} V^I_a(\phi_{0})=0$. We also have
\begin{eqnarray}
\label{Dchi}
D_\mu\chi^{I}_\nu &&=\partial_\mu \chi^{I}_\nu +\partial_\mu \phi^{J}\Gamma^{I}_{JK}\chi^{K}_\nu +A^{a}_\mu \nabla_{J}V^{I}_{a}\chi^{J}_\nu
\nonumber \\
&&=\partial_\mu \chi^{I}_\nu +\Gamma^{I}_{JK}D_\mu \phi^{J}\chi^{K}_\nu +A^{a}_\mu \partial_{J}V_{a}^{I}\chi^{J}_\nu
\notag \\
&&=\partial_\mu \chi^{I}_\nu +\Gamma^{I}_{JK}{D}_\mu \phi^{J}\chi^{K}_\nu+(A^{\vartheta a}_{0 \mu}+\tilde{A}^{a}_\mu)(\partial_{J} V_{a}^{I}+\partial_{K}\partial_{J}V_{a}^{I}\varphi^{K}+\cdots)\chi^{J}_\nu,
\end{eqnarray}
where $D_\mu \phi^J$ is as given in (\ref{Dphi}). Similarly, one can compute the expansion of $D_\mu\eta^{\bar{I}}$.

 Because of (\ref{Dphi}) and (\ref{Dchi}), we can rewrite our Lagrangian as
\begin{equation}
L=\textrm{quadratic part\ +\ vertices},
\end{equation}
where
\begin{eqnarray}
\notag
&&\textrm{quadratic part}=
k\left(\begin{array}{ccc}\varphi^{I} & \varphi^{\bar{I}} & \tilde{A}^{a}_{\mu}\end{array}\right)
             \cdot L_{\rm boson} \cdot
             \left(\begin{array}{c}\varphi^{J} \\ \varphi^{\bar{J}} \\ \tilde{A}^{b}_{\rho} \end{array}\right)  +
             k(\begin{array}{cccc}\eta^{\bar{I}} & \chi^{I}_{\mu} & \bar{c}^{a} & c^{a}\end{array})
             \cdot L_{\rm fermion} \cdot
             \left(\begin{array}{c} \eta^{\bar{J}} \\ \chi^{J}_{\rho} \\ \bar{c}^{b} \\ c^{b} \end{array}\right),
 \\
 \\  \notag
&& \textrm{vertices}=
\frac{k_{cs}}{3}\epsilon^{\mu\nu\rho}f_{abc}\tilde{A}^{a}_{\mu}\tilde{A}^{b}_{\nu}\tilde{A}^{c}_{\rho}\\ \notag
&&\ \ \ \ \ \ \ \ -\frac{k}{2}
g_{I\bar{K}}\left(\Gamma^{\bar{K}}_{\bar{M}\bar{N}}\partial_{\mu}\varphi^{\bar{M}}\chi^{I\mu}\eta^{\bar{N}}
+\nabla_{\bar{P}}V^{\bar{K}a}\eta^{\bar{P}}\chi^{I}_{\mu}\tilde{A}^{\mu}_{a}\right) \\ \nonumber
&&\ \ \ \ \ \ \ \ +\frac{k}{2\sqrt{h}}\Omega_{IJ}\epsilon^{\mu\nu\rho}\left(\Gamma^{J}_{MN}\partial_{\nu}\varphi^{M}\chi^{I}_{\mu}\chi^{N}_{\rho}
+\nabla_{P}V^{Ja}\chi^{I}_{\mu}\chi^{P}_{\rho}\tilde{A}_{\nu a}\right) \\
&&\ \ \ \ \ \ \ \ +\frac{k}{6\sqrt{h}}\Omega_{IJ}\epsilon^{\mu\nu\rho}R^{J}_{KL\bar{M}}\chi^{I}_{\mu}\chi^{K}_{\nu}\chi^{L}_{\rho}\eta^{\bar{M}}
+f_{abd}\left(\bar{c}^{a}\partial^{\mu}\tilde{A}_{\mu}^{b}c^{d}+\bar{c}^{a}\tilde{A}_{\mu}^{b}\partial^{\mu}c^{d} \right) \\
&&\ \ \ \ \ \ \ \ - \frac{k}{2}\partial_{K}\Gamma_{I\bar{M}\bar{N}}\partial_{\mu}\varphi^{\bar{M}}\varphi^{K}\chi^{I, \mu}\eta^{\bar{N}}
+\cdots . \notag
\end{eqnarray}
Here,
\begin{equation}
 L_{\rm boson}=\left( \begin{array}{ccc}
0 & 
L^{\varphi \varphi}_{I\bar{J}} & -\frac{1}{2}V_{Ib}\partial^{\rho}+A_{0 b}^{\vartheta \rho}V_{Ka}\partial_{I}V^{Kb}  \\
L^{\varphi \varphi}_{\bar{I}J}
 & 0 & -\frac{1}{2}V_{\bar{I}b}\partial^{\rho}+ A_{0 b}^{\vartheta \rho}V_{\bar{K}a} \partial_{\bar{I}}V^{\bar{K}b} \\
\frac{1}{2}V_{Ja}\partial^{\mu}+A_{0 b}^{\vartheta \mu}V_{Ka}\partial_{J}V^{Kb} & \,\,\,\, \frac{1}{2}V_{\bar{J}a}\partial^{\mu}+ A_{0 b}^{\vartheta \mu}V_{\bar{K}a} \partial_{\bar{J}}V^{\bar{K}b}  & \,\,\,\, V_{Ka}V^{K}_{b}\delta^{\mu \rho}+\frac{k_{cs}}{k}\epsilon^{\mu\nu\rho}(\kappa_{ab}\partial_{\nu}+\frac{1}{3}f_{adb}A_{0\nu}^{\vartheta d}) \end{array} \right),   \notag
\end{equation}
\begin{equation}
\label{lb}
\end{equation}
where
\begin{eqnarray}
\notag
L^{\varphi \varphi}_{I\bar{J}}:=\frac{1}{2}(g_{I\bar{J}}\partial^{2}+g_{K\bar{L}}A_{0\mu}^{\vartheta a}A_{0}^{\vartheta \mu b}\partial_{I}V^{K}_{a}\partial_{\bar{J}}V^{\bar{L}}_{b})
+g_{P\bar{J}}\partial_{I}V^{P}_{a}A^{\vartheta a}_{0\mu}\partial^{\mu},\\
L^{\varphi \varphi}_{\bar{I}J}:=\frac{1}{2}(g_{\bar{I}J}\partial^{2}+g_{K\bar{L}}A_{0\mu}^{\vartheta a}A_{0}^{\vartheta \mu b}\partial_{J}V^{K}_{a}\partial_{\bar{I}}V^{\bar{L}}_{b})+g_{J\bar{P}}\partial_{\bar{I}}V^{\bar{P}}_{a}A^{\vartheta a}_{0\mu}\partial^{\mu};
\end{eqnarray}
\begin{equation}
\label{lf}
L_{\rm fermion}=\left( \begin{array}{cccc}
0 & -\frac{1}{2}(g_{\bar{I}J}\partial^{\rho}+g_{J\bar{K}}\partial_{\bar{I}}V^{\bar{K}}_{a}A^{\vartheta a\rho}_{0}) & 0 & 0\\
\frac{1}{2}(g_{I\bar{J}}\partial^{\mu}+g_{I\bar{K}}\partial_{\bar{J}}V^{\bar{K}a}A_{0a}^{\vartheta \mu}) & \,\,\,\, \frac{\Omega_{IJ}}{2\sqrt{h}}\epsilon^{\mu\nu\rho}\partial_{\nu}+
\frac{\Omega_{IK}}{2\sqrt{h}}\epsilon^{\mu\nu\rho}\partial_{J}V^{K}_{a}A_{0\nu}^{\vartheta a} & \,\,\,\, -\frac{1}{2}\partial_{I}{\mu_{+b}}\partial^{\mu} & 0 \\
0 & \frac{1}{2}\partial_{J}{\mu_{+a}}\partial^{\rho} & 0 & -\frac{1}{2}\delta^{a}_{b}\partial^{2} \\
0 & 0 & \frac{1}{2}\delta^{a}_{b}\partial^{2} & 0 \end{array} \right);
\end{equation}
the terms $g_{IK}$ and $\Gamma^{\bar{K}}_{\bar{M}\bar{N}}$ are evaluated at $\phi_{0}$; and $\cdots$ are other terms expanded around $\phi_{0}$. Also notice that we ignore the multiplier field $B$ in $L_{\rm boson}$, for it would just be integrated out to give the gauge-fixing condition $f^{a}=0$, where we have chosen the gauge
\begin{equation}
f^{a}=\partial^{\mu}A_{\mu}^{a}=0.
\end{equation}

We can further separate the integration over the fermion zero modes $\eta_0$ and $\chi_0$ in the path-integral and write\footnote{Here, in addition to footnote~\ref{H1M}, we assume that $H^0(M, E) = 0$ so that there are no zero modes for the ghost fields $c$, $\bar c$ and $B$.}
\be
\label{par}
\hspace{-0.5cm}Z = k^{2n}\sum_{A^{\vartheta}_{0}} \left( e^{-\int_{M}L_{cs}(A^{\vartheta}_{0})}
\int_{\mathcal{M^\vartheta}}
\prod_{I=1}^{2n} d\phi^{I}_{0}
\prod_{\bar I=1}^{2n} d\phi^{\bar{I}}_{0}
 \prod_{\bar I_i=1}^{2nb^{'}_{0}}\prod_{I_j=1}^{2n b^{'}_{1}} \int d\eta_0^{\bar{I}_{i}} \int d\chi^{I_{j}}_{0 \mu} \int D\varphi D\tilde\chi D \tilde\eta D\tilde{A} Dc  D{\bar c} \, e^{-S_{A^\vartheta_{0},\phi_{0}}} \right),
\ee
where $b^{'}_{0}$ and $b^{'}_{1}$ denote the number of fermionic zero modes $\eta^{\bar I}_0$ and $\chi^I_0$, respectively; $\tilde \eta$ and $\tilde \chi$ are the corresponding \emph{nonzero} modes; and $k^{2n}$ is the normalization factor carried by the bosonic zero modes. One should note that the fermionic zero modes $\eta^{\bar I}_0$ and $\chi^I_0$ are no longer harmonic forms on $M$ like in RW theory; this is because in our case, the kinetic operator of the fermionic fields $L_{\rm fermion}$ in (\ref{lf}) is no longer the Laplacian operator but a covariant version thereof. In the limit $A \to 0$, $b^{'}_{0},\ b^{'}_{1}$ become the respective Betti numbers of $M$, while (\ref{par}) becomes the partition function of the RW theory.

\newsubsection{One-Loop Contribution }

As usual, the one-loop contribution to the perturbative partition function is given by
\begin{equation}
\label{1-loop}
Z_{0}=\int D\varphi D\tilde{\chi} D\tilde{\eta} Dc D\bar{c} \ e^{-S_{0}},
\end{equation}
where $S_{0}$ is quadratic in the fluctuating bosonic fields $\{ \tilde{A}^{\mu}_{a}, \varphi^{i}(x) \}$ and the fermionic {\it nonzero} modes $\{ \tilde \eta^{I}, \tilde \chi^{I}_{\mu} \}$:
\begin{equation}
\label{S0}
S_{0}= \int_{M} k(\begin{array}{ccc}
\varphi^{I} & \varphi^{\bar{I}} & \tilde{A}^{a}_{\mu} \end{array})
\cdot L_{\rm boson} \cdot
\left(\begin{array}{c}\varphi^{J} \\ \varphi^{\bar{J}} \\ \tilde{A}^{b}_{\rho} \end{array}\right)
+
k(\begin{array}{cccc} \tilde\eta^{\bar{I}} & \tilde \chi^{I}_{\mu} & \bar{c}^{a} & c^{a}\end{array})
\cdot L_{\rm fermion} \cdot
\left(\begin{array}{c} \tilde \eta^{\bar{J}} \\ \tilde \chi^{J}_{\rho} \\ \bar{c}^{b} \\ c^{b} \end{array}\right).
\end{equation}
Here, the tensors $g_{I\bar{J}}, \Omega_{IJ}$ and $\Gamma^{I}_{JK}$ which appear in $ L_{\rm boson}$ and $L_{\rm fermion}$ are evaluated at some $\phi_{0}$ in $\mathcal{M^\vartheta}$.

To compute (\ref{1-loop}), we first diagonalize $L_{\rm boson}$ and $L_{\rm fermion}$:
\begin{eqnarray}
\label{dia}
P_{B}^{T} \cdot L_{\rm boson} \cdot P_{B}=L^{'}_{\rm boson}, \\ \notag 
P_{F}^{T} \cdot L_{\rm fermion} \cdot P_{F}=L^{'}_{\rm fermion},
\end{eqnarray}
where $L^{'}_{\rm boson}$ and $L^{'}_{\rm fermion}$ are diagonal matrices, and $P_{B}$ and $\ P_{F}$ are orthonormal matrices ($P^{T}=P^{-1}$) constructed from the eigenvectors of $L_{\rm boson}$ and $L_{\rm fermion}$. Because $P P^{T}=1$, we can rewrite $S_{0}$ as
\begin{equation}
S_{0}^{'}= \int_{M} k(\begin{array}{ccc}
\varphi^{'I} & \varphi^{'\bar{I}} & \tilde{A}^{'a}_{\mu} \end{array})
\cdot L_{\rm boson}^{'} \cdot
\left(\begin{array}{c}\varphi^{'J} \\ \varphi^{'\bar{J}} \\ \tilde{A}^{'b}_{\rho} \end{array}\right)
+
k(\begin{array}{cccc} \tilde\eta^{'\bar{I}} & \tilde \chi^{'I}_{\mu} & \bar{c}^{'a} & c^{'a}\end{array})
\cdot L_{\rm fermion}^{'} \cdot
\left(\begin{array}{c} \tilde \eta^{'\bar{J}} \\ \tilde \chi^{'J}_{\rho} \\ \bar{c}^{'b} \\ c^{'b} \end{array}\right),
\end{equation}
where 
\begin{equation}
\left(\begin{array}{c}\varphi^{'J} \\ \varphi^{'\bar{J}} \\ \tilde{A}^{'b}_{\rho} \end{array}\right)
:=
P^{T}_{B}
\left(\begin{array}{c}\varphi^{J} \\ \varphi^{\bar{J}} \\ \tilde{A}^{b}_{\rho} \end{array}\right),
\end{equation} 
and
\begin{equation}
\left(\begin{array}{c} \tilde \eta^{'\bar{J}} \\ \tilde \chi^{'J}_{\rho} \\ \bar{c}^{'b} \\ c^{'b} \end{array}\right)
:=
P^{T}_{F}
\left(\begin{array}{c} \tilde \eta^{\bar{J}} \\ \tilde \chi^{J}_{\rho} \\ \bar{c}^{b} \\ c^{b} \end{array}\right).
\end{equation}

Moreover, the Jocabian determinants 
\begin{equation}
{\rm det}(P)={\rm det} (P^{T})=1
\end{equation}
 for both the bosonic and fermionic fields. Therefore, the measure of the path integral is such that 
\begin{equation}
D\varphi D\tilde{\chi} D\tilde{\eta} Dc D\bar{c} = D\varphi^{'} D\tilde{\chi}^{'} D\tilde{\eta}^{'} Dc^{'} D\bar{c}^{'}.
\end{equation}

In all, this means that the one-loop partition function can be rewritten as 
\begin{equation}
Z_{0}=\int D\varphi^{'} D\tilde{\chi}^{'} D\tilde{\eta}^{'}Dc^{'}D\bar{c}^{'} e^{S_{0}^{'}}.
\end{equation}

Now because $L_{\rm boson}^{'}$ and $L_{\rm fermion}^{'}$ are diagonal matrices, the path integral becomes a Gaussian integral which can be directly computed as
\begin{equation}
Z_{0}=\left(\frac{{\rm det}^{''}L_{\rm fermion}^{'}}{{\rm det}^{''}L_{\rm boson}^{'}}\right)^{\frac{1}{2}},
\end{equation}
where by using (\ref{dia}) and ${\rm det}(P^{T}P)=1$, we finally get
\begin{equation}
\label{Z0}
Z_{0}=\left(\frac{\det^{''}L_{\rm fermion}}{\det^{''}{L_{\rm boson}}}\right)^{\frac{1}{2}}.
\end{equation}
Here, the superscript $''$ indicates that only nonzero modes are considered, and $L_{\rm fermion}$ and $L_{\rm boson}$ are explicitly given by (\ref{lf}) and (\ref{lb}), respectively.

As discussed in \cite{Scott},  the (magnitude of the) one-loop contribution to the perturbative partition function of CS theory on $M$ corresponds to the analytic Ray-Singer torsion of the flat connection on $M$, while the (magnitude of the) one-loop contribution to the perturbative partition function of RW theory on $M$ corresponds to the analytic Ray-Singer torsion of the trivial connection on $M$. Since our theory is a combination of both these theories, (the magnitude of) $Z_0$ ought to be related to a hybrid of these aforementioned topological invariants of $M$.

\newsubsection{The Vacuum Expectation Value of Fermionic Zero Modes}

Notice that we may call the zero modes $\chi^{I}_{0\mu}$ and $\eta_{0}^{\bar{I}}$ of the covariant Laplacian operator ${L_{\rm fermion}}$, covariant harmonic one- and zero-forms on $M$ with values in the tangent and complex-conjugate tangent fibres $V_{\phi_{0}(x)}$ and $\bar{V}_{\phi_{0}(x)}$ over ${\cal M}^\vartheta$ evaluated at the covariantly constant map $\phi_{0}(x)$. Because
\begin{eqnarray}
&&\#({\rm{zero\ modes\ of}}\ \eta^{\bar{I}})=2n\times b^{'}_{0} ,\\ \notag
&&\#({\rm{zero\ modes\ of}}\ \chi^{I}_{\mu})=2n\times b^{'}_{1},
\end{eqnarray}
only a product of $2nb^{'}_{0}$ fields $\eta^{I}_{0}$ with $2nb^{'}_{1}$ fields $\chi^{I}_{0\mu}$ has a nonzero vacuum expectation value.

Notice also that the self-products of $\eta^{\bar{I}}_{0}$ and $\chi^{I}_{0\mu}$ are elements of the space
\begin{eqnarray}
&&H_{\eta}=\wedge^{\rm max}({\Omega}^{'0}(M)\otimes \bar{V}_{\phi_{0}(x)}), \\ \notag
&&H_{\chi}=\wedge^{\rm max}({\Omega}^{'1}(M)\otimes V_{\phi_{0}(x)}),
\end{eqnarray}
where ${\Omega}^{'i}(M)$ is the space of covariant harmonic $i$-forms on $M$. There is a lattice inside ${\Omega}^{'1}(M)$ which is formed by covariant harmonic one-forms with integer-valued
integrals over dual one-cycles in $M$; let $\omega^{(\alpha)}_{\mu}$, where $1 \leq \alpha \leq b_{1}^{'}$,  be a basis of this lattice. Then, a natural measure for the fermion zero modes can be defined by normalizing the fermionic vacuum expectation values as
\be
\label{zero m}
\langle \eta_{0}^{\bar{I}_{1}}(x_{1})\cdots \eta_{0}^{\bar{I}_{2nb'_{0}}}(x_{2nb^{'}_{0}}) \rangle = 
{k^{-nb^{'}_{0}}}\epsilon^{\bar{I}_{1}\cdots \bar{I}_{2nb^{'}_{0}}}:=
 \frac{k^{-nb^{'}_{0}}}{2n}\sum_{s\in S_{2nb_{0}^{'}}}(-1)^{|s|}\epsilon^{\bar{I}_{s(1)}\bar{I}_{s(2)}}\cdots\epsilon^{\bar{I}_{s(2nb^{'}_{0}-1)}\bar{I}_{s(2nb^{'}_{0})}},
\ee
and
\begin{eqnarray}
\label{zero m1}
\langle \chi^{I_{1}}_{0\mu_{1}}(x_{1})\cdots \chi^{I_{2nb'_{1}}}_{0\mu_{2nb^{'}_{1}}}(x_{2nb^{'}_{1}}) \rangle & = & \frac{k^{-nb^{'}_{1}}}{((2n)!)^{b^{'}_{1}}}\sum_{s\in S_{2nb^{'}_{1}}}(-1)^{|s|}  \times \nonumber \\ && \prod^{b_{1}^{'}-1}_{\alpha=0}\left(\epsilon^{I_{s(2\alpha n+1)}\cdots I_{s(2\alpha n+2n)}}\omega^{(\alpha)}_{\mu_{s(2\alpha n+1)}}(x_{s(2\alpha n+1)})\cdots  \omega^{(\alpha)}_{\mu_{s(2\alpha n+2n)}}(x_{s(2\alpha n+2n)}) \right), \notag \\
\end{eqnarray}
say in the Feynman diagrams associated with the computation of the perturbative partition function, where $S_{m}$ is the symmetric group of $m$ elements, and $|s|$ is the parity of a permutation $s$. 

Analogous to RW theory, a choice of an overall sign in (3.31) and (3.32) for the fermionic expectation values, is equivalent to a choice of orientations on the spaces 
\begin{equation}
\label{spa}
\Omega^{'0}(M) \otimes \bar{V}_{\phi_{0}(x)}, \ \ \
\Omega^{'1}(M) \otimes V_{\phi_{0}(x)}.
\end{equation} 
As a result, the whole partition function $Z$ is an invariant of $M$ up to a choice of orientation on the
spaces (\ref{spa}), as the sign of $Z$ depends on this choice.

Note that the orientations of the spaces $\bar{V}_{\phi_{0}(x)}$ and  $V_{\phi_{0}(x)}$ are determined by the $n$th power of the two-forms $\epsilon_{\bar{I}\bar{J}}$ and $\epsilon_{IJ}$ on $\mathcal{M^{\vartheta}}$, respectively. On the other hand, since $\bar{V}_{\phi_{0}(x)}$ and  $V_{\phi_{0}(x)}$ are both even-dimensional,
the orientation on the spaces (\ref{spa}) does not depend on the choice of orientation on the spaces $\Omega^{'0}(M)$ and $\Omega^{'1}(M)$, and this is why the sign of the expectation value (\ref{zero m1}) does not depend on the choice of covariant harmonic one-forms $\omega_{\mu}^{(\alpha)}$. Therefore, the choice of orientation of the spaces (\ref{spa}) and consequently, the choice of the sign in (\ref{zero m}) and (\ref{zero m1}), can always be reduced to a canonical orientation.

In discussing this orientation dependency, we have followed the analysis in~\cite{LR-Witten}. This is because in the spaces (\ref{spa}), $\Omega^{'0}(M)$, $\Omega^{'1}(M)$ and $\mathcal{M^{\vartheta}}$ (the base space for the fibres $\bar{V}_{\phi_{0}(x)}$ and  $V_{\phi_{0}(x)}$),  are just covariant versions of the harmonic forms and space of constant bosonic maps considered in RW theory, whence the analysis would be the same.

\newsubsection{Feynman Diagrams}

Let us now analyze the Feynman diagrams associated with the computation of the perturbative partition function. Note that all diagrams which contribute to the partition function should have (i) the right number of fermionic zero modes in the corresponding vertices to absorb those that appear in the path integral measure; (ii) a $k^{-2n}$ factor for canceling the normalization factor $k^{2n}$ that accompanies the partition function in (\ref{par}), because the partition function should be independent of the coupling constant $k$.

In RW theory~\cite{LR-Witten}, only a finite number of diagrams contribute to the partition function after (i) and (ii) are satisfied. In our case however, because we have, in our action, a Chern-Simons part with coupling constant $k_{cs}\neq k$, there would be an infinite number of diagrams contributing to our partition function. Fortunately though, the analysis is still tractable whence we would be able to derive some very insightful and concrete formulas in the end, as we shall see.

\bigskip\noindent{\it Canceling the Normalization Factor of $k^{2n}$}

At any rate, before we proceed to say more about the Feynman diagrams, let us discuss how one can cancel the aforementioned normalization factor of $k^{2n}$. To this end, first note that in the CS part of the action, the gauge field has quadratic term
\begin{equation}
k_{cs}AD^{0}A=k_{cs}\epsilon^{\mu\nu\rho} A^{a}_{\mu}(\kappa_{ab}\partial_{\rho}+\frac{1}{3}f_{adb}A_{0\rho}^{\vartheta d})A^{b}_{\nu}.
\label{DA}
\end{equation}
Therefore, the propagator of the gauge field is \emph{a priori}
\begin{equation}
\triangle^{A_\mu A_\nu}\sim \frac{1}{k_{cs}}.
\end{equation}
However, upon expanding the Lagrangian around $A_{0}$ and $\phi_{0}$, the gauge field will acquire a mass term
\begin{equation}
kV_{Ka}V_{b}^{K}A^{\nu a}A_{\nu }^{b}.
\end{equation}
As such, the propagator would become
\begin{equation}
\label{prop}
\triangle^{A_\mu A_\nu}=(k_{cs}D^{0}+kV_{K}\cdot V^{K})^{-1}.
\end{equation}
That being said, because the partition function does not depend on $k$, we can choose
\begin{equation}
\frac{k_{cs}}{k}\gg 1.
\label{limit2}
\end{equation}
In turn, this means from (\ref{prop}) that
\begin{equation}
\triangle^{A_{\mu}A_{\nu}}\sim \frac{1}{k_{cs}}.
\end{equation}
Hence, in what follows, we will note that $\triangle^{A_{\mu}A_{\nu}}\sim k_{cs}^{-1}$, while the other propagators are $\sim k^{-1}$.

Now, let us consider a diagram with $V$ vertices, emanating $L$ legs. Assume that this diagram contains $V_{cs}$ vertices $\frac{k_{cs}}{3}A\wedge A\wedge A$ which therefore contribute a factor of $k_{cs}^{V_{cs}}$; all the other $V-V_{cs}$ vertices therefore contribute a factor of $k^{V-V_{cs}}$. Let $L_{cs}$ be the total number of legs which are joined together by the propagator $\triangle^{A_{\mu}A_{\nu}}$, where $\mu\neq\nu$; they contribute a factor of $k_{cs}^{-\frac{L_{cs}}{2}}$. As the other propagators carry a factor of $k^{-1}$, while each fermionic zero mode carries a normalization factor of $k^{-\frac{1}{2}}$, the remaining $L-L_{cs}$ legs contribute a factor of $k^{-\frac{L-L_{cs}}{2}}$. Thus, this diagram contains a factor of
\begin{equation}
k^{-\left(\frac{L-L_{cs}}{2}-(V-V_{cs})\right)},
\end{equation}
but because the partition function is independent of $k$, it must be that
\begin{equation}
\label{cont}
\frac{L-L_{cs}}{2}-(V-V_{cs})=2n.
\end{equation}
In other words, our diagrams must obey (\ref{cont}) so that the normalization factor of $k^{2n}$ can be cancelled out.

Notice that in the case where $A \to 0$ whence $L_{cs} = V_{cs} = 0$ and our model reduces to the RW model, (\ref{cont}) would coincide with \cite[eqn.~(3.25)]{LR-Witten}, as expected. 

\bigskip\noindent{\it The Structure of the Feynman Diagrams}

\begin{figure}
\scalebox{0.8}[0.8]{\includegraphics{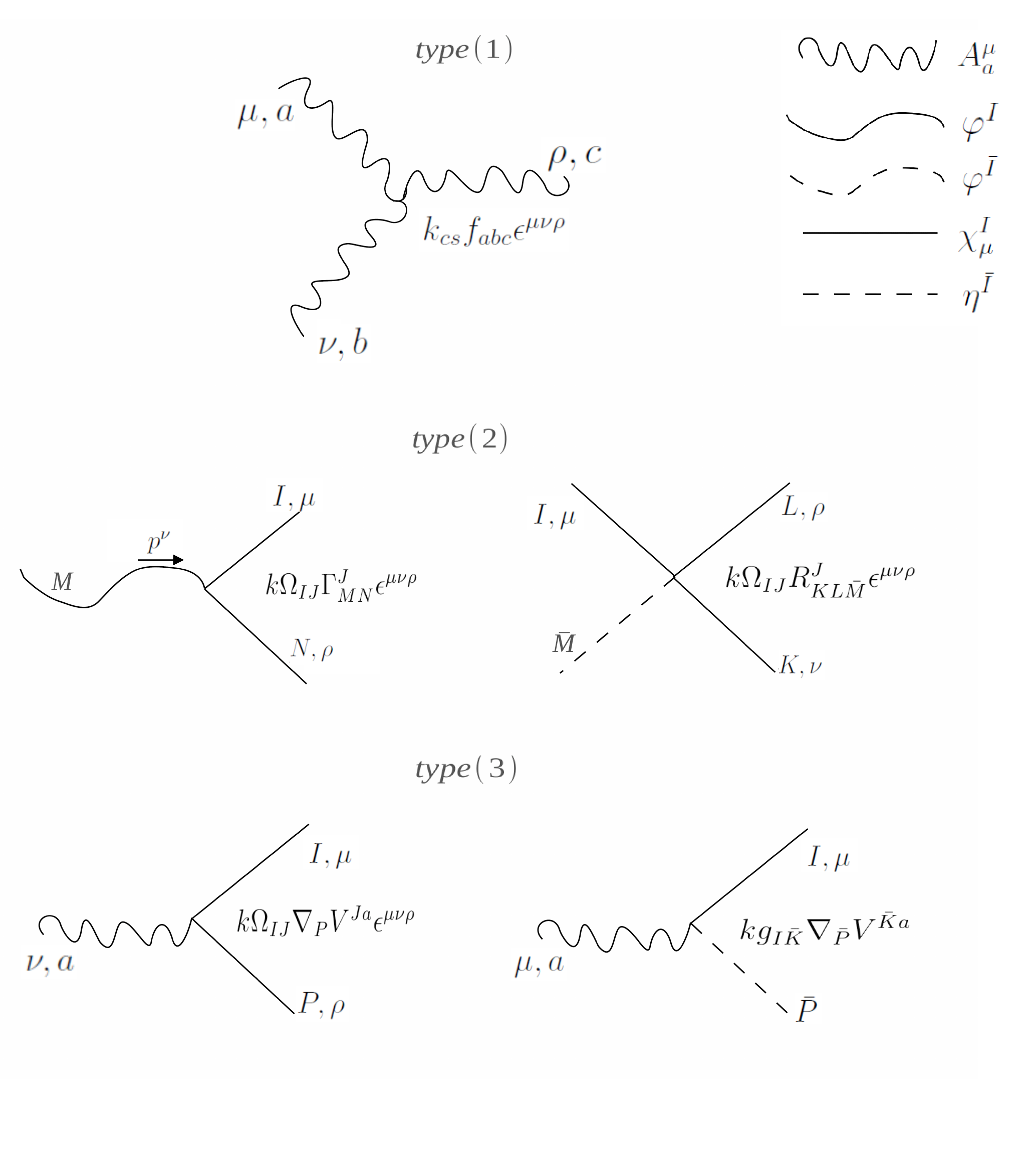}}
\caption{The Three Types of Vertices}
\end{figure}

Note that although the computation of the partition function involves summing an infinite number of Feynman diagrams because there is no constraint on $k_{cs}$, one can actually classify the vertices they involve into three types.

(1) The pure gauge field vertex coming from the CS interaction
\begin{equation}
k_{cs} f_{abc}\epsilon^{\mu\nu\rho}A^{a}_{\mu}A^{b}_{\nu}A^{c}_{\rho}.
\end{equation}

(2) The vertices free of gauge fields, such as
\begin{equation}
k\Omega_{IJ}\epsilon^{\mu\nu\rho}\Gamma^{J}_{MN}\partial_{\nu}
\varphi^{M}\chi^{I}_{\mu}\chi^{N}_{\rho} \qquad {\rm and} \qquad  k\Omega_{IJ}\epsilon^{\mu\nu\rho}R^{J}_{KL\bar{M}}\chi^{I}_{\mu}
\chi^{K}_{\nu}\chi^{L}_{\rho}\eta^{\bar{M}}.
\end{equation}

(3) The vertices that mix matter fields\footnote{Here, for convenience, we use ``matter fields'' to mean $\phi$, $\chi$ and $\eta$.} with gauge fields, such as
\begin{equation}
k\Omega_{IJ}\epsilon^{\mu\nu\rho}\nabla_{P}V^{Ja}\chi^{I}_{\mu}\chi^{P}_{\rho}A_{\nu a} \qquad {\rm and} \qquad
k g_{I\bar{K}}\nabla_{\bar{P}}V^{\bar{K} a}\eta^{\bar{P}}\chi^{I}_{\mu}A^{\mu}_a.
\end{equation}
These three types of vertices are illustrated in Fig.~1.

\newsubsection{The Propagator Matrices and an Equivariant Linking Number of Knots}

In order to compute the Feynman diagrams,  one would also need to have a knowledge of the propagators of the bosonic and fermionic fields associated with the kinetic operators  $L_{\rm boson}$ and $L_{\rm fermion}$.

The propagator of the bosonic fields $\triangle_{\rm boson}$ can be obtained by solving the equation
\begin{equation}
kL_{{\rm boson}(IK,\bar{I}\bar{K}, ad, \mu\rho)}(x)\times \triangle^{(KJ,\bar{K}\bar{J}, db, \rho\nu)}_{\rm boson}(x-y)=\left( \begin{array}{ccc}
\delta^{I}_{J}&0&0\\
0&\delta^{\bar{I}}_{\bar{J}}&0\\
0&0&\delta^{\mu}_{\nu}\delta^{a}_{b}
\end{array} \right)
\cdot \delta(x-y),
\end{equation}
where $L_{\rm boson}$ is given in (\ref{lb}). To first order, the $3\times 3$ matrix $\triangle_{\rm boson}$ can be written as
\begin{equation}
\triangle_{\rm boson} \sim \left( \begin{array}{ccc}
0&\triangle^{\varphi\varphi}&\triangle^{\varphi A}\\
\triangle^{\varphi\varphi}&0&\triangle^{\varphi A}\\
\triangle^{A\varphi}&\triangle^{A\varphi}&\triangle^{AA}
\end{array} \right),
\end{equation}
where its components are spanned by all possible boson propagators:
\begin{eqnarray}
\label{boson propogator}
\notag
&&\triangle^{(\varphi\varphi)I\bar{J}}(X, G, M;\phi_{0}, A^\vartheta_{0})=\frac{1}{k}f^{(\varphi\varphi)I\bar{J}}(X, G;\phi_{0},A^\vartheta_{0})\triangle^{'(\varphi\varphi)}(M;A^\vartheta_{0}),\\
&&\triangle^{(AA)ab}_{\mu\nu}(X, G, M;\phi_{0}, A^\vartheta_{0})=\frac{1}{k_{cs}}f^{(AA)ab}(X, G;\phi_{0},A^\vartheta_{0})\triangle^{'(AA)}_{\mu\nu}(M;A^\vartheta_{0})  , \\
&&\triangle^{(\varphi A)I,a}_{\mu}(X, G, M;\phi_{0}, A^\vartheta_{0})=\frac{1}{k}f^{(\varphi A)I,a}(X, G;\phi_{0},A^\vartheta_{0})\triangle^{'(\varphi A)}_{\mu}(M;A^\vartheta_{0}). \notag
\end{eqnarray}
Here, the labels $\phi_0$ and $A^\vartheta_{0}$ mean that the corresponding quantities are evaluated at these values of the covariantly constant map $\phi_0$ and flat connection $A^\vartheta_{0}$. Notice that we can write the propagators as a product of two parts. The first part is a function $f(X, G;\phi_{0},A^\vartheta_{0})$ on the target manifold $X$ that is characterized by the structural information of $X$ and $G$. The second part  is a function $\triangle^{'}(M;A^\vartheta_{0})$ on $M$.

Similarly, the propagator of the fermionic fields $\triangle_{\rm fermion}$ can be obtained by solving the equation
\begin{equation}
kL_{{\rm fermion}(\bar{I}\bar{K}, IK, ad, \mu\rho)}(x)\times \triangle^{(\bar{K}\bar{J}, KJ, db, \rho\nu)}_{\rm fermion}(x-y)=\left( \begin{array}{cccc}
\delta^{\bar{I}}_{\bar{J}}&0&0&0\\
0&\delta^{I}_{J}\delta^{\mu}_{\nu}&0&0\\
0&0&\delta^{a}_{b}&0\\
0&0&0&\delta^{a}_{b} \end{array} \right)
\cdot
\delta(x-y),
\end{equation}
where $L_{\rm fermion}$ is given in (\ref{lf}). To first order, the $4\times 4$ matrix $\triangle_{\rm fermion}$ can be written as
\begin{equation}
\triangle_{\rm fermion} \sim \left( \begin{array}{cccc}
0&\triangle^{\eta\chi}&0&0\\
\triangle^{\chi\eta}&\triangle^{\chi\chi}&\triangle^{\chi \bar{c}}&0\\
0&\triangle^{\bar{c}\chi}&0&\triangle^{c\bar{c}}\\
0&0&\triangle^{\bar{c}c}&0
\end{array} \right).
\end{equation}
where its components are spanned by all possible fermion propagators:
\begin{eqnarray}
\label{fermion propogator}
\notag
&&\triangle^{(\eta\chi)\bar{I}J}_{\mu}(X, G, M;\phi_{0}, A^\vartheta_{0})=\frac{1}{k}f^{(\eta\chi)\bar{I}J}(X, G;\phi_{0},A^\vartheta_{0})\triangle^{'(\eta\chi)}_{\mu}(M;A^\vartheta_{0})   , \\ \notag
&&\triangle^{(\chi\chi)IJ}_{\mu\nu}(X, G, M;\phi_{0}, A^\vartheta_{0})=\frac{1}{k}f^{(\chi\chi)IJ}(X, G;\phi_{0},A^\vartheta_{0})\triangle^{'(\chi\chi)}_{\mu\nu}(M;A^\vartheta_{0}) ,       \\
&&\triangle^{(\chi\bar{c})I,a}_{\mu}(X, G, M;\phi_{0}, A^\vartheta_{0})=\frac{1}{k}f^{(\chi\bar{c})I,a}(X, G;\phi_{0},A^\vartheta_{0})\triangle^{'(\chi\bar{c})}_{\mu}(M;A^\vartheta_{0})     ,   \\   \notag
&&\triangle^{({c}\bar{c})ab}(X, G, M;\phi_{0}, A^\vartheta_{0})=\frac{1}{k}f^{({c}\bar{c})ab}(X, G;\phi_{0},A^\vartheta_{0})\triangle^{'(c\bar{c})}(M;A^\vartheta_{0}). \notag
\end{eqnarray}
Similar to the boson propagators, we can also write these fermion propagators as the product of two parts.

\bigskip\noindent{\it  An Equivariant Linking Number of Knots}

Notice here that we may regard $\Delta^{'(\chi\chi)}_{\mu\nu}(M;A^\vartheta_{0})$ as an equivariant one-form depending on $A^\vartheta_{0}$. This means that for one-cycles $C^{'}$ in $M$ which satisfy the following equivariant Stoke's theorem
\begin{equation}
\int_{C^{'}}d_{A^\vartheta_0}\mathcal{F}=\int_{\partial_{A^\vartheta_{0}}C^{'}}\mathcal{F}=0,
\end{equation}
the double integral
\begin{equation}
\oint_{C^{'}_{1}}dx^{\mu}\oint_{C^{'}_{2}}dx^{\nu}\Delta^{'(\chi\chi)}_{\mu\nu}(M;A^\vartheta_{0})
\end{equation}
would define an ``\emph{equivariant linking number}'' of knots $C^{'}_{1}$ and $C^{'}_{2}$.

\newsubsection{New Three-Manifold Invariants and Weight Systems}

We would now like to show that by computing the perturbative partition function, we would be able to derive \emph{new} three-manifold invariants and their associated weight systems which depend on both $G$ and $X$. To this end, let us first review the three-manifold invariants and their associated weight systems that come from Chern-Simons and Rozansky-Witten theory.

\bigskip\noindent{\it  Three-Manifold Invariants and Weight Systems From Chern-Simons Theory}

The perturbative partition function of Chern-Simons theory can  be written as
\begin{equation}
Z_{CS}(M;G; k_{cs})=\sum_{m}Z_{CS}^{(m)}(M; G; k_{cs}),
\end{equation}
where $(m)$ denotes the order of $k_{cs}$ in the indicated term. If the classical solution $A_{0}$ is the trivial flat connection over $M$, the propagators would be independent of $A_{0}$. Then, the partition function would take  (up to a one-loop contribution) the very simple form
\begin{equation}
Z^{(\rm tr)}_{CS}(M;G; k_{cs}) = \exp\left(\sum_{m=1}^{\infty}S_{G, m+1}(M) \, k_{cs}^{-m}\right),
\end{equation}
where
\begin{equation}
S_{G, m+1}=\sum_{\Gamma \in \Gamma_{3, m+1}}a_{\Gamma}(G)I_{\Gamma}(M).
\label{weight}
\end{equation}
Here, the sum runs over all trivalent Feynman graphs $ \Gamma_{3, m+1}$ with $m+1$ loops (and $2m$ vertices),\footnote{For a description of a (trivalent) Feynman graph, see~\cite{LR-Witten}.} and $I_{\Gamma}(M)$ are the integrals over $M \times M \times \dots \times M$ of the products of propagators.

The Jacobi identity of the Lie algebra of $G$ is used to show that although the individual integrals $I_{\Gamma}(M)$ depend on the metric of $M$, the metric-dependence cancels out of the sum in (\ref{weight})~\cite{Scott, E. Guad}. Thus, $S_{G, m+1}$ and therefore  $Z_{CS}(M; G; k_{cs})$, are indeed topological invariants of the three-manifold $M$. Furthermore, because the factor $a_{\Gamma}(G)$ can be regarded as a weight factor weighting each graph term,  $S_{G, m+1}$ also defines what is called a weight system. Clearly, this weight system depends on Lie algebra structure.

\bigskip\noindent{\it  Three-Manifold Invariants and Weight Systems From Rozansky-Witten Theory}

The perturbative partition function of Rozansky-Witten theory can (up to a one-loop contribution) be written as
\begin{equation}
\label{Z-RW}
Z(M, X)=\sum_{\Gamma}Z_{\Gamma}(M, X),
\end{equation}
where $\sum_{\Gamma}$ is a summation over all relevant Feynman graphs of the theory, and
\begin{equation}
\label{ws-RW}
Z_{\Gamma}(M, X)= b_{\Gamma}(X)\sum_{b}I_{\Gamma, b}(M),
\end{equation}
where $\sum_{b}$ denotes the summation of all possible ways of assigning the vertices to each Feynman graph. Here, $I_{\Gamma, b}$ are the integrals over $M \times M \times \dots \times M$ of the products of propagators as well as of the relevant one-form fermionic zero modes. $I_{\Gamma, b}$ just depends on the structure of $M$, while $b_{\Gamma}$ serves as a weight factor which depends on the curvature tensor of the target space $X$ that comes from the underlying vertices. Thus, $Z_{\Gamma}(M, X)$ defines a weight system. Clearly, this weight system depends on hyperk\"ahler geometry. 

The Bianchi identity plays the same role here as the Jacobi identity in CS theory~\cite{LR-Witten}; one can use it to show that the dependence on the metric of $M$ cancels out of the sum (\ref{Z-RW}), i.e., $Z(M,X)$ is a topological invariant of the three-manifold $M$.

\bigskip\noindent{\it Coming Back to Our Theory}

Coming back to our theory, we can, after evaluating the path integral, write the perturbative partition function as
\begin{equation}
Z(M, X, G)=\sum_{A^\vartheta_0} \, e^{-\int_{M}k_{cs}L_{cs}(A^{\vartheta}_{0})}\cdot Z_0 (A^\vartheta_0) \cdot Z(M, X, G; A^{\vartheta}_{0}; k_{cs}),
\label{Z expan}
\end{equation}
where $e^{-\int_{M}k_{cs}L_{cs}(A^{\vartheta}_{0})}$ is the topological factor coming from  the Chern-Simons part of the total Lagrangian evaluated at a flat connection $A_{0}^{\vartheta}$; $Z_0(A^\vartheta_0)$ is the topological one-loop contribution given in (\ref{Z0}); and
 \begin{equation}
 \label{Z-sum}
 Z(M, X, G;A_{0}^{\vartheta}; k_{cs})= \sum_{\Gamma} {Z_{\Gamma}}(M, X, G; A_{0}^{\vartheta}; k^m_{cs}),
 \end{equation}
where $\sum_{\Gamma}$ is a sum over all possible Feynman diagrams with two or more loops that (i) have the right number of fermionic zero modes to absorb those that appear in the path integral measure, and (ii) are free of the coupling constant $k$. Here, the label $k^m_{cs}$ (where $m$ may vanish) means that $\Gamma$ carries with it a factor of $k^m_{cs}$.

In fact, $Z_{\Gamma}$ can be expressed as
\begin{equation}
 Z_{\Gamma}(M, X, G; A^{\vartheta}_{0}; k^m_{cs})=\int_{ \mathcal{M^\vartheta}}\sqrt{g}\ d^{2n}\phi^{I}_{0}\ d^{2n}\phi^{\bar{I}}_{0}
\, \,  W_{\Gamma}(X, G;\phi_{0},A^{\varphi}_{0})  I_{\Gamma}(M, X, G;\phi_{0}, A_{0}^{\vartheta}; k^m_{cs}),
                                 \label{parti}
\end{equation}
where $I_{\Gamma}$ is an integral over $M \times M \times \dots \times M$ of the products of propagators as well as of the one-form fermionic zero modes $\omega_\mu(x)$ in (\ref{zero m1}), while the weight factor $W_{\Gamma}$ is a product of terms relevant to $\Gamma$ that are associated with the vertices in Fig.~1.

We can characterize the partition function by classifying the Feynman diagrams into three categories as follows. \newline

(1) \underline{Chern-Simons-Type Diagrams}. These diagrams result purely from the vertices $A\wedge A\wedge A$. Thus, they correspond to diagrams in usual Chern-Simons theory. The topological property of Chern-Simons-type diagrams has already been verified in earlier works~\cite{Scott, E. Guad}. As such, we would have nothing more to add about them. \newline

To discuss the next two types of diagrams, we take, for simplicity, the case where $b^{'}_{1}=0$ and $b^{'}_{0}=1$. Then, the nonvanishing Feynman diagrams must contain exactly $2nb_{0}^{'}$ zero modes $\eta^{I}_0$. For brevity, we will only discuss diagrams whose vertices emanate 4 legs.  \newline

\begin{figure}
\scalebox{0.7}[0.7]{\includegraphics{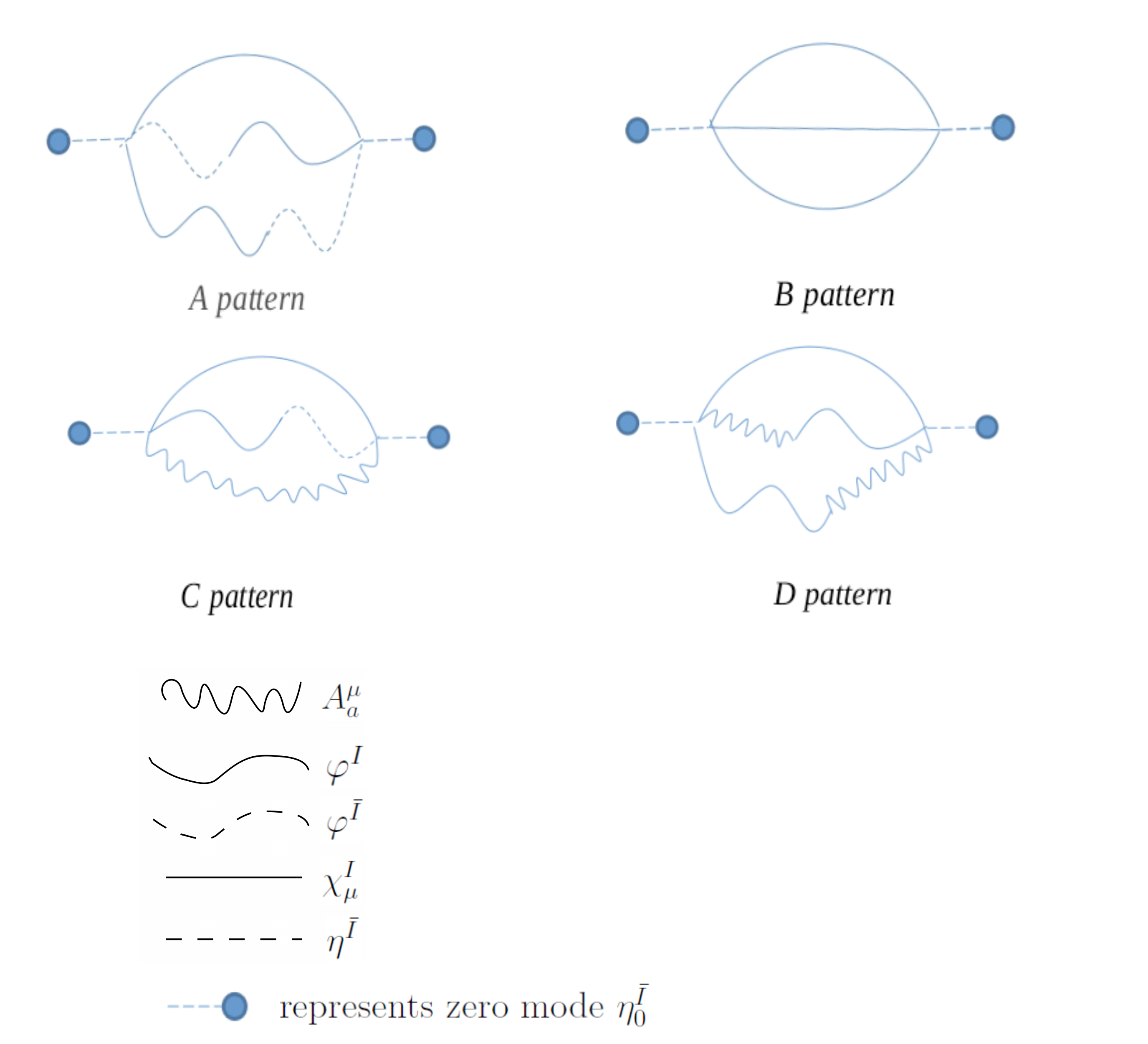}}
\caption{$A$, $B$, $C$ and $D$ Pattern Diagrams}
\end{figure}

(2) \underline{Diagrams Free of Gauge Fields}. Since these diagrams result from vertices which are free of the gauge field $A$,  they do not contain the gauge field propagator. Examples of such diagrams are  given by the $A$ pattern and $B$ pattern in Fig.~2.

The $A$ pattern diagram is formed by the vertex
\begin{equation}
k\partial_{K}\Gamma_{I\bar{M}\bar{N}}\partial_{\mu}\varphi^{\bar{M}}\varphi^{K}\chi^{I, \mu}\eta_{0}^{\bar{N}}, \notag
\end{equation}
and the propagators in the diagram are $\triangle^{(\chi\chi)}$ and $\triangle^{(\varphi\varphi)}$. Therefore for the $A$ pattern diagram,  the terms in (\ref{parti}) are
\begin{equation}
\boxed{ I_{\Gamma}(M, X, G;\phi_{0}, A^{\vartheta}_{0})=\int_{M}\prod_{l=1}^{n} (\triangle^{(\chi\chi)I_{l}J_{l}}_{\mu\nu}
                                        \partial^{\mu}\triangle^{(\varphi\varphi)L_{l}\bar{P}_{l}}
                                        \partial^{\nu}\triangle^{(\varphi\varphi)K_{l}\bar{Q}_{l}}) (x_l, y_l) \, d^3x_l d^3 y_l}
\end{equation}
and
\begin{equation}
\boxed{W_{\Gamma}(X, G;\phi_{0}^{I})=\epsilon^{\bar{M}_{1}\dots \bar{M}_{2n}}
                                 \prod_{l=1}^{n}(\partial_{K_{l}}\partial_{\bar{M}_{1, l}}g_{\bar{P}_{l}I_{l}})
                                                 (\partial_{{L}_{l}}{\partial_{\bar{M}_{2, l}}}g_{\bar{Q}_{l}J_{l}}) }
\end{equation}
where the $\epsilon^{\bar{M}_{1}\dots \bar{M}_{2n}}$ factor comes from the expectation value of the zero modes $\eta_{0}$ defined in (\ref{zero m}). The contribution of this diagram to the partition function can then be evaluated by substituting the above two expressions in (\ref{parti}).

The $B$ pattern diagram is formed by the vertex
\begin{equation}
k\Omega_{IJ}R^{J}_{KL\bar{M}}\epsilon^{\mu\nu\rho}\chi^{I}_{\mu}\chi^{K}_{\nu}\chi^{L}_{\rho}\eta_{0}^{\bar{M}}, \notag
\end{equation}
and the propagator in the diagram is $\triangle^{(\chi\chi)}$. Therefore for the $B$ pattern diagram,  the terms in (\ref{parti}) are
\begin{equation}
\boxed{I_{\Gamma}(M, G, X;\phi_{0}^{I}, A^{\vartheta}_{0})=\int_{M}\prod_{l=1}^{n} (\epsilon^{\mu_{1}\nu_{1}\rho_{1}}\epsilon^{\mu_{2}\nu_{2}\rho_{2}}
                                       \triangle^{(\chi\chi)I_{1, l}I_{2, l}}_{\mu_{1}\mu_{2}}
                                       \triangle^{(\chi\chi)K_{1, l}K_{2, l}}_{\nu_{1}\nu_{2}}
                                       \triangle^{(\chi\chi)L_{1, l}L_{2, l}}_{\rho_{1}\rho_{2}}) (x_l, y_l) \, d^3x_l d^3 y_l  }
\end{equation}
and
\begin{equation}
\boxed{W_{\Gamma}(X, G;\varphi_{0}^{I})=\epsilon^{\bar{M}_{1}\dots \bar{M}_{2n}}
                                 \prod_{l=1}^{n}{(\Omega_{I_{1, l}J_{1, l}}R^{J_{1, l}}_{K_{1, l}L_{1, l}\bar{M}_{1, l}})
                                                 (\Omega_{I_{2, l}J_{2, l}}R^{J_{2, l}}_{K_{2, l}L_{2, l}\bar{M}_{2, l}})}}
\end{equation}

Notice that the $A$ and $B$ pattern diagrams in Fig.~2 are similar to those in RW theory. Nevertheless, unlike RW theory, our propagator factor $I_{\Gamma}$ depends on the flat gauge field $A^\vartheta_{0}$. If $A^\vartheta_{0}$ were trivial, the contributions of the $A$ and $B$ pattern diagrams to the partition function would be as given in RW theory, as expected. \newline

(3) \underline{Diagrams Mixing Gauge and Matter Fields}. Examples of such diagrams are given by the $C$ pattern and $D$ pattern in Fig.~2.

The $C$ pattern diagram is formed by the vertices
\begin{equation}
k\partial_{K}(\nabla_{\bar{M}}V^{a}_{I})\eta_{0}^{\bar{M}}\chi^{I}_{\mu}A^{\mu}_{a}\varphi^{K}\quad {\rm{and}} \quad     k\partial_{\bar{K}}(\nabla_{\bar{M}}V^{a}_{I})\eta_{0}^{\bar{M}}\chi^{I}_{\mu}A^{\mu}_{a}\varphi^{\bar{K}}, \notag
\end{equation}
and the propagators in the diagram are $\triangle^{(\chi\chi)}$,  $\triangle^{(\varphi\varphi)}$,  $\triangle^{(AA)}$. Therefore for the $C$ pattern diagram,  the terms in (\ref{parti}) are
\begin{equation}
\boxed{I_{\Gamma}(M, X, G;\phi^{I}_{0}, A^{\vartheta}_{0}; k^{-n}_{cs})=  \int_{M}\prod_{l=1}^{n}
                    (\triangle^{(\chi\chi)I_{l}J_{l}}_{\mu\nu}
                    \triangle^{(\varphi\varphi)K_{l}\bar{L}_{l}}
                    \triangle^{(AA)\mu\nu}_{a_{l}b_{l}}
) (x_l, y_l) \, d^3x_l d^3 y_l}
\end{equation}
and
\begin{equation}
\boxed{W_{\Gamma}(X, G;\varphi_{0}^{I})=\epsilon^{\bar{M}_{1}\dots \bar{M}_{2n}}
                                 \prod_{l=1}^{n}{\partial_{\bar{L}_{l}}{(\nabla_{\bar{M}_{1, l}}V_{I_{l}}^{a_{l}})}
                                                 \partial_{K_{l}}{(\nabla_{\bar{M}_{2, l}}V_{J_{l}}^{b_{l}})}}}
\end{equation}

The $D$ pattern diagram is formed by the vertex
\begin{equation}
k\partial_{K}(\nabla_{\bar{M}}V^{a}_{I})\eta_{0}^{\bar{M}}\chi^{I}_{\mu}A^{\mu}_{a}\varphi^{K}, \notag
\end{equation}
and the propagators in the diagram are $\triangle^{(\chi\chi)}$,  $\triangle^{(\varphi A)}$. Therefore for the $D$ pattern diagram,  the terms in (\ref{parti}) are

\begin{equation}
\boxed{I_{\Gamma}(M, X, G;\phi^{I}_{0}, A^{\vartheta}_{0})=\int_{M}\prod_{l=1}^{n}
                                       (\triangle^{(\varphi A)K_{1, l}}_{\mu a_{l}}
                                       \triangle^{(\varphi A)K_{2, l}}_{\nu b_{l}}
                                       \triangle^{(\chi\chi)I_{1, l}I_{2, l}, \mu\nu}) (x_l, y_l) \, d^3x_l d^3 y_l }
\end{equation}
and
\begin{equation}
\boxed{ W_{\Gamma}(X, G;\varphi_{0}^{I})=\epsilon^{\bar{M}_{1}\dots \bar{M}_{2n}}
                                 \prod_{l=1}^{n}{\partial_{K_{1, l}}{(\nabla_{\bar{M}_{1, l}}V_{I_{1, l}}^{a_{l}})}
                                                 \partial_{K_{2, l}}{(\nabla_{\bar{M}_{2, l}}V_{I_{2, l}}^{b_{l}})}}  }
\end{equation}

\bigskip\noindent{\it New Three-Manifold Invariants and Weight Systems}

As shown in (\ref{boson propogator}) and (\ref{fermion propogator}),  the propagators $\triangle$ can be expressed as the product of a function $f(X, G)$ on the target manifold $X$ and a function $\triangle^{'}(M)$ on the three-manifold $M$. Therefore, we can rewrite the above propagator factors as
\begin{equation}
\boxed{I_{\Gamma}(M, X, G;\phi_{0}, A^{\vartheta}_{0}; k^m_{cs})=f_{\Gamma}(X, G;\phi_{0}, A^{\vartheta}_{0})I^{'}_{\Gamma}(M;A^{\vartheta}_{0}; k^m_{cs})}
\end{equation}
where $I^{'}_{\Gamma}$ is a function on $M$ that depends on the flat gauge field $A^{\vartheta}_{0}$ and which carries a factor of $k^m_{cs}$, and the function $f_\Gamma$ is characterized, among other things, by the structure of the target space $X$ and the gauge group $G$. In turn, this means that we can rewrite (\ref{parti}) as
\begin{equation}
\label{par2}
{
 Z_{\Gamma}(M, X, G; A^{\vartheta}_{0}; k^m_{cs})=  \mathscr W_{\Gamma} (X, G; A^{\vartheta}_{0}) \, I^{'}_{\Gamma}(M;A^{\vartheta}_{0}; k^m_{cs})
                                  }
                                 \end{equation}
where
\begin{equation}
\label{W}
\boxed{\mathscr W_{\Gamma} (X, G; A^{\vartheta}_{0}) = \int_{\mathcal{M^\vartheta}} \, \sqrt{g} \, d^{2n}\phi^{I}_{0} \, d^{2n}\phi^{\bar{I}}_{0}
                                 \,\, W_{\Gamma}(X, G;\phi_{0}, A^{\vartheta}_{0})f_{\Gamma}(X, G;\phi_{0}, A^{\vartheta}_{0})}
\end{equation}
can be regarded as a weight factor which combines the structural information of the hyperk\"ahler manifold $X$ and the Lie algebra  $\mathfrak{g}$ of the gauge group $G$.

As in CS and RW theory, $I^{'}_{\Gamma}$ in (\ref{par2}) can be expected to depend on the metric of $M$. However,  since the partition function in (\ref{Z expan}) and therefore
 \begin{equation}
 \label{Z-sum-2}
 \boxed{Z(M, X, G;A_{0}^{\vartheta}; k_{cs})= \sum_{\Gamma}  \mathscr W_{\Gamma} (X, G; A^{\vartheta}_{0}) \, I^{'}_{\Gamma}(M;A^{\vartheta}_{0}; k^m_{cs})}
 \end{equation}
are topological on $M$ at the outset, the metric-dependence of $I^{'}_{\Gamma}$ should cancel out in the sum (\ref{Z-sum-2}). To rigorously show this cancellation, we can use the Jacobi identity of $G$,  the Bianchi identity of $X$, and the geometric identities of the moment maps discussed in section $2.1$. However, at each order of $k_{cs}$, the partition function and consequently, its variation with respect to the metric of $M$, contains so many different terms that it would be a formidable task to demonstrate this cancellation using our purely physical methods. We hope that in the near future, novel and sophisticated methods would be devised to facilitate this explicit verification.

 In summary, our perturbative partition function furnishes us with a \emph{new} three-manifold invariant  $Z(M, X, G;A_{0}^{\vartheta}; k_{cs})$ which depends on both $G$ and $X$, that also defines a \emph{new} weight system whose weights $\mathscr W_{\Gamma} (X, G; A^{\vartheta}_{0})$ are characterized by both Lie algebra structure \emph{and} hyperk\"ahler geometry.

\newsection{Canonical Quantization and the Nonperturbative Partition Function}

\newsubsection{Canonical Quantization}

Let us now canonically quantize our gauged sigma model on $M$ with target space $X$. To this end, first note that the Hamiltonian $H= \langle \int T_{00} \rangle = \langle \delta_{\hat{Q}}\mathcal{O} \rangle=0$. Next, note that locally, the Riemannian manifold $M$ can be written as $\Sigma\times I$,  where $I$ is the `time' dimension and $\Sigma$ is a compact Riemann surface. Thus, since $H=0$ whence the theory should be time-independent, it would mean that we can just analyze the physics over any $\Sigma \times I \subset M$. 

This property of $H =0$ also means that only ground states contribute to the spectrum of the theory. Therefore, where the fermions are concerned, only the zero modes contribute to the physical Hilbert space. Where the gauge field is concerned, only the classical configuration of flat connections $A_{0}$ contribute to the physical Hilbert space. And where the bosons are concerned, only the covariantly constant maps $\phi$ from $M$ to $X$ which satisfy $D_{\mu}\phi=\partial_{\mu}\phi+A^a_{0 \mu}V_a=0$, contribute to the physical Hilbert space.  

Let $\tau$ be the time coordinate. Then, according to the last paragraph, $\phi$ would satisfy $\partial_{\tau}\phi+A^a_{0 \tau}V_a=0$. In the gauge where $A_{\tau}=0$, we would also have
\begin{equation}
\partial_{\tau}\eta^{\bar{I}}=0, \quad \partial_{\tau}\chi_{\tau}^{I}=0, \quad \partial_{\tau} \chi^I_\mu = 0,
\end{equation}
where $\eta^{\bar I}$, $\chi^{I}_\tau$ and $\chi^I_\mu$ are fermionic zero modes. In other words, the zero modes $\phi$, $\eta^{I}$, $\chi_{\tau}^{I}$ and $\chi^I_\mu$  are $\tau$-independent, which means that we effectively have a two-dimensional gauged sigma model on $\Sigma$.

\bigskip\noindent{\it The Commutation and Anticommutation Relations}

From the Lagrangian, we compute the momentum conjugate of $\eta$, $\chi$ and $A$ to be
\begin{eqnarray}
\label{cc}
\frac{\delta L}{\delta\partial^{\tau}\eta^{\bar{K}}}=g_{I\bar{K}}\chi_{\tau}^{I}, \notag \\ 
\frac{\delta L}{\delta\partial_{\tau}\chi_{\mu}^{I}}=\epsilon^{\tau\mu\nu}\Omega_{IJ}\chi_{\nu}^{J}, \notag \\
\frac{\delta L}{\delta\partial^{\tau}\chi_{\tau}^{I}}=g_{\bar{K}I}\eta^{\bar{K}},\\
\frac{\delta L}{\delta\partial_{\tau}A_{0\mu}^{a}}=\epsilon^{\tau\mu\nu}k_{ab}A_{0\nu}^{b}. \notag
\end{eqnarray}
Note at this point from (\ref{lf}) that the fermionic zero modes $\chi_\mu^I$ are solutions of the covariant equation
\begin{equation}
\label{def}
\Omega_{IJ}\epsilon^{\tau\mu\nu}\partial_{\mu}\chi_{\nu}^{J}
+\Omega_{IK}\epsilon^{\tau\mu\nu}\partial_{J}V_{a}^{K}A_{0\mu}^{a}\chi_{\nu}^{J}=0; \quad \mu,\nu = 1, 2,
\end{equation}
which depend on the choice of the flat connection $A_{0}$; in other words, we can write $\chi_{\mu}^{I}(x)=\chi_{\alpha}^{I}\omega_{\mu}^{\alpha}(A_{0}, x)$, where $\omega^\alpha$ are covariant harmonic one-forms on $\Sigma$, and $\chi_{\alpha}^{I}$ are constant fermionic coefficients. Hence, if $\int_{\Sigma} \omega^{\alpha}\wedge \omega^{\beta} = L^{\alpha, \beta}$, the relations in (\ref{cc}) tell us that the commutation and anticommutation relations upon quantizing the zero modes must be 
\begin{eqnarray}
\label{cc-ac}
\notag
\{\eta^{\bar{I}}, \chi_{\tau}^{J}\}=\frac{1}{k}g^{\bar{I}J},\\
\{\chi_{\alpha}^{I}, \chi_{\beta}^{J}\}=\frac{1}{k}\Omega^{IJ}(L^{-1})_{\alpha, \beta}, \\
\left[A_{0\mu}^{a}(x)A_{0\nu}^{b}(y)\right]=\frac{1}{k_{cs}}\epsilon_{\mu\nu}\delta^{ab}\delta^{2}(x-y), \notag
\end{eqnarray}
where $g$ and $\Omega$ are evaluated at the covariantly constant map $\phi$.

\bigskip\noindent {\it A Relevant Digression}

Before proceeding any further, let us discuss the following important point. Recall from section 3 that after gauge-fixing, it is the $\hat Q$-cohomology that is relevant. Nevertheless, the spectrum of the theory is unchanged by gauge-fixing, and so the $Q$- and $\hat Q$-cohomology ought to be equivalent. Let us now verify this claim. 

First, recall that we have
\be
Q^2 = \textrm{gauge transformation},
\ee
and
\begin{equation}
\hat{Q}=Q+Q_{\rm FP} , \quad \textrm{where} \quad \hat{Q}^{2}=0. 
\end{equation}
 
Second, by definition, we have
\be
\label{ker}
{\rm ker}(\hat{Q}) =  \{\mathcal{O} |  \{Q+Q_{\rm FP}, \mathcal{O} ] =0\}. 
\ee
From (\ref{BRSTgf}), we find that $\{ Q, \mathcal{O} ] \neq -  \{ Q_{\rm FP}, \mathcal{O}]$.  So,
\begin{equation}
 \label{kerr}
 {\rm ker}(\hat{Q})={\rm ker}(Q)\cap {\rm ker}(Q_{\rm FP}).
\end{equation}
That is, $ \{Q, \mathcal{O}]=0$ and $ \{Q_{\rm FP}, \mathcal{O} ] = 0$, which means 
\begin{equation}
\{ \{Q, Q_{\rm FP}\}, \mathcal{O} ]=0 .
\label{qqf}
\end{equation}

Third, from the definition of $Q_{\rm FP}$, we have
\begin{equation}
\{ Q_{\rm FP}^2, \mathcal{O}]=0.
\label{qfp}
\end{equation}
Also, we have
\begin{equation}
\{\hat{Q}^2, \mathcal{O} ] =  \{ Q^2+ \{Q, Q_{\rm FP}\}+ Q_{\rm FP}^{2}, \mathcal{O} ]=0.
\label{hatq}
\end{equation}
Thus, from (\ref{qqf}), (\ref{qfp}) and (\ref{hatq}), we have
\begin{equation}
 \{Q^{2}, \mathcal{O} ] = 0.
\end{equation}
In turn, this means that
\begin{equation}
{\rm ker}(\hat{Q})={\rm ker}(Q)\cap {\rm ker}(Q_{\rm FP})= {\rm ker}(Q)\cap \{\mathcal{O}| \{ Q^{2}, \mathcal{O} ]=0 \}.
\end{equation}

Last, note that in
\begin{equation}
{\rm im}(\hat{Q})= {\rm im} (Q+Q_{\rm FP}), 
\end{equation}
because ${\rm im} (Q_{\rm FP})$ contains ghost fields, ${\rm im} (Q_{\rm FP})$ does not contribute to the physical Hilbert space. Hence,
\begin{equation}
{\rm im}(\hat{Q})\cong {\rm im}(Q).
\end{equation}

Altogether, this means that
\begin{equation}
\mathcal{O}\in \frac{{\rm ker}(\hat{Q})}{{\rm im}(\hat{Q})}=\frac{{\rm ker}(Q)\cap {\rm ker}(Q_{\rm FP})}{{\rm im}(Q+Q_{\rm FP})}\cong \frac{{\rm ker}(Q)\cap \{\mathcal{O}|{ \{ Q^{2}, \mathcal{O} ]=0 }\}}{{\rm im}(Q)},
\end{equation}
which verifies our claim that the $\hat Q$- and $Q$-cohomology are equivalent.  Therefore, let us henceforth focus on the $Q$-cohomology; in particular, let us proceed to ascertain the relevant Hilbert space of states in the $Q$-cohomology.

\bigskip\noindent{\it The Hilbert Space of States}

To this end, note that since we are restricting ourselves to the classical configuration $A_0$ that is free of interacting fluctuations, we can view the total theory as a CS theory plus a non-dynamically gauged RW theory. As such, any state $| \Psi \rangle$ in the  $Q$-cohomology ought to take the form
\begin{equation}
\boxed{|\Psi \rangle = | \psi \rangle \otimes \tilde{\Phi}|0 \rangle}
\label{Hil}
\end{equation}
Here,  $| \psi \rangle$ is a state in the CS theory which is associated with a $Q$-closed but not $Q$-exact wave function $\psi(A_{0}^{\vartheta})$  that depends on a flat gauge field $A_{0}^{\vartheta}$ along $\Sigma$, where~\cite{Marcos}
\be
\label{psi-path}
\psi(A_{0}^{\vartheta})=\int_{A|_{\Sigma} = A_{0}^{\vartheta}} DA \, e^{-S_{cs}},
\ee
and $\tilde{\Phi}$ is a $Q$-closed but not $Q$-exact state operator of the non-dynamically gauged RW theory. Let us now determine  $\tilde{\Phi}$.

From (\ref{cc-ac}), it is clear that the vacuum state $|0 \rangle$ would be annihilated by the operators $\chi^{I}_{\beta}$ and $\chi^{I}_{\tau}$.\footnote{For ease of illustration, we henceforth assume that $b^{'}_{0}=1$ and $b^{'}_{1}=1$.}
Hence, a first-cut construction of an arbitrary state $|\Phi \rangle$ of the non-dynamically gauged RW theory would be
\begin{equation}
|\Phi \rangle= \tilde \Phi | 0 \rangle = \Phi_{I_{1}\cdots I_{l} \bar{I_{1}}\cdots\bar{I}_{k}} (\phi)
\chi^{I_{1}}_{\alpha}\cdots \chi^{I_{l}}_{\alpha}
\eta^{\bar{I}_{1}}\cdots\eta^{\bar{I}_{k}}|0 \rangle.
\end{equation}
Generically, $\Phi$ depends on the covariantly constant map $\phi$; hence, a natural generically-nonvanishing scalar product of states would be given by
\be
\langle \Phi^{(1)} |\Phi^{(2)} \rangle = \int_{{\cal M}^\vartheta}  {\sqrt g} \, d^{2n} \phi^I d^{2n} \phi^{\bar I} \, \epsilon^{I_1 \dots I_{2n} \bar I_1 \dots \bar I_{2n}} \, \Phi^{(1)}_{I_{1}\cdots I_{q} \bar{I_{1}}\cdots\bar{I}_{p}}  \Phi^{(2)}_{I_{q+1}\cdots I_{2n} {\bar I}_{p+1} \cdots \bar{I}_{2n}} =  \int_{{\cal M}^\vartheta} \Phi^{q+ p} \wedge \Phi^{4n - q - p}, 
\ee  
where $\Phi^m$ is an $m$-form on ${\cal M}^\vartheta$, the space of all physically distinct $\phi$'s for some $A^\vartheta_0$. In other words, $\tilde \Phi$ would correspond to an element of $\Omega^{l+k}({\cal M}^\vartheta)$, the space of all $(l+k)$-forms on ${\cal M}^\vartheta$. 

Now, from (\ref{BRST}), the fields transform under the supercharge $Q$ as
\begin{eqnarray}
\label{variations}
\notag
\delta_{Q}\eta^{\bar{I}_{i}}=-V_{a}^{\bar{I}_{i}}\mu_{+}^{a},\\
\delta_{Q}\phi^{\bar{I}}=\eta^{\bar I}, \\ \notag
\delta_{Q}\chi^{I}=D\phi^{I}.
\end{eqnarray}
At the level of zero modes, the last equation $\delta_{Q}\chi^{I}=D\phi^{I}=0$. If $\tilde \Phi$ is $d$-closed, i.e., $\partial_{\bar{J}}\Phi(\phi)= \partial_{{J}}\Phi(\phi) = 0$, we would have
\begin{eqnarray}
\label{progress 1}
\delta_{Q} \tilde \Phi &&= \sum_{i}(-1)^{l+ i}\Phi_{I_{1}\cdots I_{l} \bar{I}_{1}\cdots\bar{I}_{k}} \, V_{a}^{\bar{I}_{i}}\mu_{+}^{a} \, 
\chi^{I_{1}}_{\alpha}\cdots\chi^{I_{l}}_{\alpha}
\eta^{\bar{I}_{1}}\cdots\hat{\eta}^{\bar{I}_{i}}\cdots\eta^{\bar{I}_{k}} \nonumber \\ 
&&= (-1)^l \, k \, \Phi_{I_{1}\cdots I_{l} \bar{I}_{1}\bar{I}_{2}\cdots\bar{I}_{k}} \, V_{a}^{\bar{I}_{1}}\mu_{+}^{a} \, 
\chi^{I_{1}}_{\alpha}\cdots\chi^{I_{l}}_{\alpha}
\eta^{\bar{I}_{2}}\cdots\eta^{\bar{I}_{k}} \\
&& \neq 0.\nonumber
\end{eqnarray}
Thus, the state operator $\tilde \Phi$ is not $Q$-closed, as we would like it to be. 

We can try to `improve' it to
\begin{equation}
\tilde{\Phi}=\Phi_{I_{1}\cdots I_{l}\bar{I}_{1}\cdots\bar{I}_{k}} (\phi)
\chi^{I_{1}}_{\alpha}\cdots\chi^{I_{l}}_{\alpha}
\eta^{\bar{I}_{1}}\cdots\eta^{\bar{I}_{k}} - \mu_{+}^{a}\Phi_{a I_{1}\cdots I_{l}\bar{I}_{1}
\cdots\bar{I}_{k-2}}(\phi) \chi^{I_{1}}_{\alpha}\cdots\chi^{I_{l}}_{\alpha}
\eta^{\bar{I}_{1}}\cdots\eta^{\bar{I}_{k-2}},
\label{equ coho}
\end{equation}
where now,
\begin{eqnarray}
\label{variation}
\notag
\delta_{Q}\tilde{\Phi}=
&& (-1)^l \, [k \, V_{a}^{\bar{I}_{1}} \Phi_{I_{1}\cdots I_{l}\bar{I}_{1}\bar{I}_{2}
\cdots\bar{I}_{k}}] \, \mu_{+}^{a} \,
\chi^{I_{1}}_{\alpha}\cdots\chi^{I_{l}}_{\alpha}
\eta^{\bar{I}_{2}}\cdots\eta^{\bar{I}_{k}}  \\
&&- (-1)^l \, [\partial_{\bar{K}}\Phi_{a I_{1}\cdots I_{l}\bar{I}_{1}
\cdots\bar{I}_{k-2}} ] \, \mu_{+}^{a} \,
\chi^{I_{1}}_{\alpha}\cdots\chi^{I_{l}}_{\alpha}
\eta^{\bar{K}}\eta^{\bar{I}_{1}}\cdots\eta^{\bar{I}_{k-2}}
\\
&&- (-1)^l (k-2) \mu_{+}^{a} \mu_{+}^{b} V_{b}^{\bar{I}_{1}} \Phi_{a I_{1}\cdots I_{l}\bar{I}_{1}\cdots\bar{I}_{k-2}}
\chi^{I_{1}}_{\alpha}\cdots\chi^{I_{l}}_{\alpha}
\eta^{\bar{I}_{2}}\cdots\eta^{\bar{I}_{k-2}}, \notag
\end{eqnarray}
after exploiting the fact that $\mu_+$ is holomorphic. If moreover, $\Phi_{a I_{1}\cdots I_{l}\bar{I}_{1}\cdots\bar{I}_{k-2}}
$ is anti-holomorphic and 
\be
i_a (\Phi) = d \Phi_a,
\ee
where $(i_a (\Phi))_{{\bar I}_2 \cdots I_{l}\bar{I}_{1}\bar{I}_{2} \cdots\bar{I}_{k}} = k \, V_{a}^{\bar{I}_{1}} \Phi_{I_{1}\cdots I_{l}\bar{I}_{1}\bar{I}_{2} \cdots\bar{I}_{k}}$ is a contraction with $V_a$ of $\Phi \in \Omega^{l+k}({\cal M}^\vartheta)$, and $(d \Phi_a)_{\bar K I_1 \cdots I_l \bar I_1 \cdots \bar I_{k-2}} = \partial_{\bar{K}}\Phi_{a I_{1}\cdots I_{l}\bar{I}_{1}
\cdots\bar{I}_{k-2}}$ is an exterior derivative of $\Phi_a \in \frak g^\ast \otimes \Omega^{l+k -2}({\cal M}^\vartheta)$, the second term on the RHS of (\ref{variation}) would simply cancel the first one out, i.e., we would have 
\be
\label{progress 2}
\delta_{Q}\tilde{\Phi} = (-1)^{l+1} (k-2) \mu_{+}^{a} \mu_{+}^{b} V_{b}^{\bar{I}_{1}} \Phi_{a I_{1}\cdots I_{l}\bar{I}_{1}\cdots\bar{I}_{k-2}}
\chi^{I_{1}}_{\alpha}\cdots\chi^{I_{l}}_{\alpha}
\eta^{\bar{I}_{2}}\cdots\eta^{\bar{I}_{k-2}}.
\ee

Hence, by `improving' $\tilde \Phi$ via (\ref{equ coho}), we have actually made progress: by comparing (\ref{progress 2})  and (\ref{progress 1}), it is clear that we have gone from having $k$ to $k-2$ many $\eta$ fields in the expression for $\delta_Q \tilde \Phi$. 

We can continue to `improve' $\tilde \Phi$ by adding more terms of lower order in $\eta$:
\begin{eqnarray}
\label{corrected phi}
\tilde{\Phi}& = & \Phi_{I_{1}\cdots I_{l}\bar{I}_{1}\cdots\bar{I}_{k}} (\phi)
\chi^{I_{1}}_{\alpha}\cdots\chi^{I_{l}}_{\alpha}
\eta^{\bar{I}_{1}}\cdots\eta^{\bar{I}_{k}} - \mu_{+}^{a}\Phi_{a I_{1}\cdots I_{l}\bar{I}_{1}
\cdots\bar{I}_{k-2}}(\phi) \chi^{I_{1}}_{\alpha}\cdots\chi^{I_{l}}_{\alpha}
\eta^{\bar{I}_{1}}\cdots\eta^{\bar{I}_{k-2}} \nonumber \\
&& \hspace{-0.0cm} + \mu_{+}^{a} \mu_{+}^{b} \Phi_{a b I_{1}\cdots I_{l}\bar{I}_{1}
\cdots\bar{I}_{k-4}} (\phi) \chi^{I_{1}}_{\alpha}\cdots\chi^{I_{l}}_{\alpha}
\eta^{\bar{I}_{1}}\cdots\eta^{\bar{I}_{k-4}}  \\
&&  - \mu_{+}^{a} \mu_{+}^{b} \mu_{+}^{c} \Phi_{a b c I_{1}\cdots I_{l}\bar{I}_{1}
\cdots\bar{I}_{k-6}} (\phi) \chi^{I_{1}}_{\alpha}\cdots\chi^{I_{l}}_{\alpha}
\eta^{\bar{I}_{1}}\cdots\eta^{\bar{I}_{k-6}} + \dots. \nonumber
\end{eqnarray}
Here, even-valued $k$ is such that $0 <  k \leq {\rm dim}_\mathbb C(\cal M^\vartheta)$, and $\Phi_a, \Phi_{ab}, \Phi_{abc}, \dots \in S(\frak g^\ast) \otimes \Omega ({\cal M}^\vartheta)$  are anti-holomorphic, where $S(\frak g^\ast)$ is the symmetric algebra on $\frak g^\ast$. If moreover, 
\be
\label{contract}
i_a (\Phi) = d \Phi_a, \quad i_b (\Phi_a) = d \Phi_{ab}, \quad i_c(\Phi_{ab}) = d \Phi_{abc}, \dots,
\ee
one will find that $\delta_Q \tilde \Phi = 0$. 

From the field variations in (\ref{variations}) and the comment thereafter, one can see that $Q$ effectively acts on $\tilde \Phi$ as $d - \mu^a_+ i_a$. Together with (\ref{corrected phi}) and (\ref{contract}),  it would mean that for $\tilde \Phi$ to be $Q$-closed but not $Q$-exact, it must correspond to a class in the $G$-equivariant cohomology $H_G({\cal M}^\vartheta)$.

It is now clear from (\ref{Hil}) and the fact that $\tilde \Phi$ corresponds to a class in $H_G({\cal M}^\vartheta)$, that the relevant Hilbert space $\cal H$ of all states $| \Psi \rangle$ in the $Q$-cohomology can be expressed as
\begin{equation}
\label{space-cq}
\boxed{\mathcal{H}= {\cal H}_{\rm CS} (A_{0}^{\vartheta}, \Sigma) \otimes H_G({\cal M}^\vartheta)}
\end{equation}
where ${\cal H}_{\rm CS} (A_{0}^{\vartheta}, \Sigma)$ is the Hilbert space of states in CS theory associated with wave functions $\psi(A_{0}^{\vartheta})$ in the $Q$-cohomology that depend on a flat gauge field $A_{0}^{\vartheta}$ on $\Sigma$.

\bigskip\noindent{\it An Example}

Before we end this subsection, let us consider the case where $\Sigma={\bf S}^{2}$, $G$ is some arbitrary compact simple Lie group, $X=T^{*}(G/\mathbb T)$, and $\mathbb T \subset G$ is a maximal torus. For simply-connected $\Sigma={\bf S}^2$, we can go to pure gauge on $\Sigma$ whence we can regard the flat gauge field $A^\vartheta_0$ to be trivial in all directions (since $A^\vartheta_{0 \tau} = 0$ also). Consequently, ${\cal H}_{\rm CS}$ is trivial, $\chi_\mu$ and $\eta$ would become ordinary harmonic forms on $\Sigma$, and the $\phi$'s would just be constant maps whence ${\cal M}^\vartheta = X = T^{*}(G/\mathbb T)$.  Therefore, the corresponding Hilbert space would simply be
\begin{equation}
\mathcal{H}_G = H_{G}(T^{*}(G/\mathbb T)),
\end{equation}
which is the $G$-equivariant cohomology of $T^{*}(G/\mathbb T)$. In other words, via the Cartan model of equivariant cohomology, we have (c.f.~\cite{SUSY Eqv})
\begin{equation}
 \mathcal{H}_G \cong H\left( [S(\mathfrak g^\ast)\otimes \Omega(T^{*}(G/\mathbb T))]_{\textrm{$G$-invariant}}\right).
\end{equation}
Here, $H( \dots)$ is the cohomology of the complex with Cartan differential $d_G = 1 \otimes d + F^a \otimes i_a$, where $F^a$ is some $\frak g$-valued function on $T^{*}(G/\mathbb T)$ of degree two.   

From our discussion leading up to (\ref{corrected phi}), and the fact that $b_{1}({\bf S}^2)=0$ and $b_{0}({\bf S}^2)=1$ whence there are no $\chi_\mu$'s but ${\rm dim}_\mathbb C(T^{*}(G/\mathbb T))$ many $\eta$'s, we find that a generic arbitrary state in ${\cal H}_G$ would be given by 
\be
| \Psi \rangle = \tilde \Phi | 0 \rangle, 
\ee
where
\begin{eqnarray}
\label{phi-example}
\tilde{\Phi}& = & \Phi_{\bar{I}_{1}\cdots\bar{I}_{k}} (\phi)
\eta^{\bar{I}_{1}}\cdots\eta^{\bar{I}_{k}} - \mu_{+}^{a}\Phi_{a \bar{I}_{1}
\cdots\bar{I}_{k-2}}(\phi) \eta^{\bar{I}_{1}}\cdots\eta^{\bar{I}_{k-2}} \nonumber \\
&&  + \mu_{+}^{a} \mu_{+}^{b} \Phi_{a b \bar{I}_{1}
\cdots\bar{I}_{k-4}}(\phi) \eta^{\bar{I}_{1}}\cdots\eta^{\bar{I}_{k-4}}  \\
&&  - \mu_{+}^{a} \mu_{+}^{b} \mu_{+}^{c} \Phi_{a b c \bar{I}_{1}
\cdots\bar{I}_{k-6}}(\phi) \eta^{\bar{I}_{1}}\cdots\eta^{\bar{I}_{k-6}} + \dots. \nonumber 
\end{eqnarray}
Here, even-valued $k$ is such that $0 <  k \leq {\rm dim}_\mathbb C(T^{*}(G/\mathbb T))$, and $\Phi_a, \Phi_{ab}, \Phi_{abc}, \dots \in S(\frak g^\ast) \otimes \Omega (T^{*}(G/\mathbb T))$  are anti-holomorphic.

That being said, it can be shown~\cite{Brion} that ${\cal H}_G  = S(\frak t^\ast)$, where $\frak t$ is the Lie algebra of $T$. In other words, an arbitrary state in ${\cal H}_G$ ought to be given by 
\be
\boxed{| \Psi \rangle = (-1)^p \, \mu_{+}^{a_1} \mu_{+}^{a_2} \cdots \mu_{+}^{a_p} \Phi_{a_1 a_2 \dots a_p}(\phi) | 0 \rangle} 
\ee
where $1 \leq a_i \leq {\rm rank} (G)$, and $p$ is any positive integer.

Take for example $G = SU(2)$ and $X = T^{*}({\bf {CP}}^1)$, where ${\rm rank}(G) =1$ and ${\rm dim}_{\mathbb C}(X) = 2$. Then, the only state in ${\cal H}_{SU(2)}$ is
\begin{eqnarray}
| \Psi^{(1)} \rangle & = & - \mu_{+}^{1}\Phi_{1}(\phi) | 0\rangle.
\end{eqnarray}

Take as another example $G = SU(N)$ and $X = T^{*}(SU(N)/U(1)^{N-1})$, where ${\rm rank}(G) =N-1$ and ${\rm dim}_{\mathbb C}(X) = N(N-1)$.  Then, the independent states in ${\cal H}_{SU(N)}$ ought to take the form
\be
| \Psi^{(i)} \rangle  =   (-1)^i \mu_{+}^{a_1} \mu_{+}^{a_2} \cdots \mu_{+}^{a_i} \Phi_{a_1 a_2 \dots a_i} (\phi) | 0\rangle, 
 \ee
where $1 \leq  i \leq N(N-1)/2$. 

One can proceed to compute ${\cal H}_G$ for any $G$ in a similar manner. For brevity, we shall leave this to the interested reader.

\newsubsection{The Nonperturbative Partition Function}

We shall now furnish a general prescription that will allow us to compute, nonperturbatively, the partition function of our model on any  three-manifold with target space $X$.

Suppose we have manifolds $M_{1}$ and $M_{2}$ whose boundaries are the same compact Riemann surface $\Sigma$ but with opposite orientations, such that after gluing them along $\Sigma$, we get a new manifold $M$. Then, from the axioms of quantum field theory, the partition function on $M$ with target space $X$ would be given by
\begin{equation}
Z_{X}(M)= \langle M_{2}|M_{1}\rangle.
\end{equation}
Here, $|M_{1}\rangle \in {\cal H}_1$ is a state due to the path integral over $M_1$ that is associated with $\Sigma$, and $|M_{2}\rangle \in {\cal H}_2$ is a state due to the path integral over $M_2$ that is also associated with $\Sigma$, where the Hilbert spaces ${\cal H}_1$ and ${\cal H}_2$ are canonically dual to each other. 
 
We could also twist the boundary of $M_{1}$ by an element $U$ of the mapping class group of $\Sigma$ prior to gluing, whence the partition function on the resulting three-manifold $M^{U}$ would be given by
\begin{equation}
Z_{X}(M^{U})= \langle M_{2}|\hat U|M_{1} \rangle,
\end{equation}
where $\hat U$ is an operator acting in ${\cal H}_1$ that  represents $U$. In this manner, one can, with appropriate choices of $\Sigma$, $M_{1}$ and $M_2$, construct any three-manifold $M^U$, and upon determining how $\hat U$ acts on $| M_1 \rangle$ to produce another state in ${\cal H}_1$, the corresponding partition function on $M^U$ can be determined via a tractable calculation on $M$. Therefore, let us determine the action of $\hat U$ on $ |M_1 \rangle$. 

For concreteness, let us consider $\Sigma = {\bf T}^2$ whence $U$ is an element of  $SL(2, \mathbb Z)$, and $M_1$ is  a solid torus. Then, we can conveniently choose on $\Sigma$, basic one-forms $\xi^{1,2}$ and basic one-cycles $C_{1,2}$, whereby
\begin{equation}
\int_{C_{b}}\xi^{a}=\delta_{ab},
\end{equation}
so that the matrix
\begin{equation}
U=
\left( \begin{array}{cc}
p & q \\
r & s
\end{array} \right)  \in SL(2,\mathbb Z), \ \
ps-qr=1,
\end{equation}
transforms the pair of cycles $(C_{1},\ C_{2})$ as
 \begin{equation}
 U:
  \left(\begin{array}{c}
C_{1} \\
C_{2}
\end{array} \right)
\to
\left( \begin{array}{cc}
p & q \\
r & s
\end{array} \right)
\left( \begin{array}{c}
C_{1} \\
C_{2}
\end{array} \right).
\end{equation}

Now, from (\ref{Hil}), we have
\begin{equation}
|M_1 \rangle = |\psi_0\rangle \otimes \tilde{\Phi}_1|0 \rangle,
\end{equation}
where the subscripts `0' and `1' accompanying $|\psi \rangle$ and $\tilde \Phi$ are just convenient labels to associate them to $| M_1 \rangle$.

Let us first determine how $\hat U$ acts on $|\psi_0\rangle$. From the explanation of CS theory in~\cite{Witten-Jones}, since $|\psi_0\rangle$ is associated with a path integral on $M_1$ with no operator insertions (see (\ref{psi-path})),  we can regard it as a vector $v_0$ in the Verlinde basis of ${\cal H}_{\rm CS} (A_{0}^{\vartheta}, {\bf T}^2)$, the space of integrable representations of the affine algebra associated with $G$ at level $k_{cs}$, where the subscript `$0$' in $v_0$ means that it is associated with the trivial representation of $G$~\cite[section 4.3]{Witten-Jones}. As such, according to \emph{loc.~cit.}, we have
\be
\hat U  |\psi_0 \rangle = K_0{}^j | \psi_j \rangle,
\ee
where $|\psi_j \rangle$ corresponds to the vector $v_j$ in ${\cal H}_{\rm CS} (A_{0}^{\vartheta}, {\bf T}^2)$ that is associated with the $R_j$ representation of $G$; the $R_j$'s are in one-to-one correspondence with the highest weights of $G$; $K$ is the Verlinde matrix~\cite{matrix}; and the underlying wave function is
\be
\psi_j (A_{0}^{\vartheta}) = \int_{A|_{{\bf T}^2} = A_{0}^{\vartheta}} DA \, \, {W}_j(C) \, e^{-S_{cs}},
\ee
where
\be
{W}_j(C)={\rm Tr}_{R_j} P \exp\left(\oint_{C} A^a\mu_{+ a}\right)
\ee
is the $Q$-closed (and therefore gauge-invariant) trace of the holonomy of the one-form $A^a\mu_{+ a}$ along the longitudinal cycle $C$ in $M_1$ taken in the representation $R_j$.\footnote{Note that the arguments in~\cite[section 4.3]{Witten-Jones} involve the Wilson loop operator $\mathscr W_j(C) = {\rm Tr}_{R_j} P \exp\left(\oint_{C} A^aT_a \right)$ and not $W_j(C) ={\rm Tr}_{R_j} P \exp\left(\oint_{C} A^a\mu_{+ a}\right)
$, where the $T_a$'s are generators of the Lie algebra $\frak g$ of $G$.  Nevertheless, recall that $d\mu_{+ a}=-i_{V_{a}}(\Omega)$, where the vector fields $V_{a}$ associated with the $G$-action on $X$ are generators of $\frak g$; in other words, like the $T_a$'s, the $\mu_{+a}$'s can be labeled by representations of $G$. Also, under a gauge transformation with parameter $\Lambda$, we have  $\delta_{\Lambda}(\mu_{+ a})=-f^{d}_{a c}\Lambda_{d}\mu_{+}^{c}$ and $\delta_{\Lambda}T_{a}=f^{d}_{ac}\Lambda_{d}T^{c}$. Last but not least, we have the Poisson bracket relation (\ref{PB-mu}). Altogether, this means that we can, for all our purposes, regard $\mu_{+a}$ as the matrix $T_a$ whence we can also regard $W_j$ as a Wilson loop operator $\mathscr W_j$. \label{u-rep}}

Next, let us determine the action of $\hat U$ on $\tilde{\Phi}_1$, where $\tilde{\Phi}_1$ takes the generic form in (\ref{corrected phi}). As $\eta^{\bar{I}}$ is a (geometrically-trivial) scalar on $\Sigma$, we only need to consider the action on $\chi^{I}_{\alpha}$.

Recall that we can write 
\be
\chi_{\mu}^{I} =\chi_{\alpha}^{I}\omega_{\mu}^{\alpha}(A^\vartheta_{0}),
\ee
where the $\omega^\alpha$'s are covariant harmonic one-forms on $\Sigma$. This is similar to the case in RW theory, except that here, the $\omega^\alpha$'s also depend on a certain flat connection  background $A^\vartheta_{0}$. Since we are free to choose $A^\vartheta_{0}$, let us choose a background whereby there are two covariant harmonic forms on $\Sigma$, i.e., there are two solutions to (\ref{def}). Then,  the fermionic zero modes can be expressed as
\begin{equation}
\label{chi-1}
\chi^{I}_{\beta}=\int_{C^{'}_{\beta}} \chi^{I}_{\alpha}\omega^{\alpha}, \quad {\alpha, \beta = 1, 2,}
\end{equation}
since
\begin{equation}
\label{chi}
\int_{C^{'}_{\beta}}\omega^{\alpha} =\delta^{\alpha}_{\beta},
\end{equation}
where $C^{'}_{\beta}$ are a pair of covariant basic one-cycles in ${\bf T}^2$. Assuming that our background is also such that $C^{'}_{1,2}$ is not deformed away from $C_{1,2}$, we also have
 \begin{equation}
 U:
  \left(\begin{array}{c}
C_{1}^{'} \\
C_{2}^{'}
\end{array} \right)
\to
\left( \begin{array}{cc}
p & q \\
r & s
\end{array} \right)
\left( \begin{array}{c}
C_{1}^{'} \\
C_{2}^{'}
\end{array} \right).
\end{equation}
Then, according to (\ref{chi}) and (\ref{chi-1}), 
\begin{equation}
\label{chi-tx}
\hat U:
  \left(\begin{array}{c}
\chi^I_{1} \\
\chi^I_{2}
\end{array} \right)
\to
\left( \begin{array}{cc}
p & q \\
r & s
\end{array} \right)
\left( \begin{array}{c}
\chi^I_{1} \\
\chi^I_{2}
\end{array} \right),
\end{equation}
where we shall regard $\chi^I_2$ to be the annihilation operator. Therefore, 
\begin{equation}
\label{map}
\hat U: \tilde{\Phi}_1(\phi, \eta, \chi^I_{1}) \to \tilde{\Phi}_1(\phi, \eta, p\chi^I_{1}+q\chi^I_{2}).
\end{equation}
We would like to emphasize that the modular transformation will not modify the intrinsic definition of the zero modes $\chi$, $\phi$ and $\eta$ (which depends on the respective differential operators covariant with respect to $A^\vartheta_0$).\footnote{This claim can be justified as follows. First, note that $A^\vartheta_0$, being flat, is a covariant harmonic one-form on $\Sigma$, just like $\chi^I$; hence, its components $A^\vartheta_{01}$ and $A^\vartheta_{02}$ will transform as in (\ref{chi-tx}). Nevertheless, the covariant equation $\Omega_{IJ}\epsilon^{\tau\mu\nu}\partial_{\mu}\chi_{\nu}^{J} +\Omega_{IK}\epsilon^{\tau\mu\nu}\partial_{J}V_{a}^{K}A_{0\mu}^{a}\chi_{\nu}^{J}=0; \, \, \mu,\nu = 1, 2$, which defines $\chi$, is a scalar equation on $\Sigma$ (since the $\mu$ and $\nu$ indices are fully contracted) -- it is thus insensitive to the transformation of $\Sigma$ by $U$ whence the definition of $\chi$ would be unmodified. As for $\phi$ and $\eta$, they are defined by the following one-form equations on $\Sigma$ (since there is a free $\mu$ index): $\partial_\mu \phi + A^a_\mu V_a = 0$ and $g_{I\bar{J}}\partial^{\mu}\eta^{\bar{J}}+g_{I\bar{K}}\partial_{\bar{J}}V_{a}^{\bar{K}}A_{0}^{a\mu}\eta^{\bar{J}}=0$. If we rewrite these equations as $D_\mu \phi = 0$ and ${\cal D}^\mu \eta = 0$, $\mu = 1, 2$, then the action of $U$ on $\Sigma$ would map the first equation from ${D_1 \phi = 0} \to { p D_1 \phi + q D_2 \phi = 0}$ and ${D_2 \phi = 0} \to { r D_1 \phi + s D_2 \phi = 0}$. But $D_1 \phi$ and $D_2 \phi$ are independent quantities whence $ p D_1 \phi + q D_2 \phi = 0$ and  $r D_1 \phi + s D_2 \phi = 0$ imply that $D_\mu \phi = 0$, $\mu = 1, 2$, which is the same as the original equation. The same argument applies for the second equation involving $\eta$. Hence, the definition of $\phi$ and $\eta$ would also be unmodified.}  Hence, the (generic) intrinsic definition (\ref{corrected phi})  of $\tilde \Phi_1$ will not be modified either. Thus, there is no ambiguity in the map (\ref{map}). 

At any rate, the property that the vacuum $| 0 \rangle$ would be annihilated by $\chi^I_2$ must also hold after a transformation by $\hat U$; in other words, if
\be
\chi^I_2 |0 \rangle = 0, 
\ee
then
\be
\hat U: | 0 \rangle \to | 0 \rangle',  
\ee
where
\be
\label{rchi}
(r\chi_{1}^{I}+ s\chi_{2}^{I}) | 0 \rangle' = 0.
\ee
Since (\ref{cc-ac}) means that we can represent $\chi^I_1$ by multiplication and $\chi^I_2$ by $\Omega^{IJ} \partial/ \partial \chi_1^J$,  we can also write (\ref{rchi}) as
\be
\left(s \Omega^{IJ} {\partial \over \partial \chi_1^J} +  r \chi_{1}^{I} \right) | 0 \rangle' = 0, 
\ee 
which implies that
\be
|0\rangle' =    {\rm exp} \left( - {r \over s} \Omega_{IJ} \chi^I_1 \chi^J_1 \right)| 0 \rangle.
\ee

In all, this means that $\hat U$ acts on $| M_1 \rangle$ as
\be
\hat U: | M_1 \rangle \to | M_1 \rangle',
\ee 
where
\be
| M_1 \rangle' =    K_0{}^j |\psi_j \rangle  \otimes \tilde{\Phi}_1\left(\phi, \eta, p\chi_{1}^{I}+q\Omega^{IJ} {\partial \over \partial \chi_1^J} \right)  \cdot {\rm exp} \left( - {r \over s} \Omega_{IJ} \chi^I_1 \chi^J_1 \right) |0\rangle.
\ee

We are now ready to compute the partition function $Z_{X}(M^{U})$. If
\be
\langle M_2 | = \langle 0 | \tilde{\Phi}^\dagger_2(\phi, \eta, \chi_{1}^{I}) \otimes \langle \gamma|,
\ee
 then $Z_{X}(M^{U})= \langle M_{2}| \hat U |M_{1} \rangle =   \langle M_{2}|M_{1} \rangle'$ can also be expressed as
\be
Z_{X}(M^{U})=   K_0{}^j \langle \gamma |\psi_j \rangle \cdot \langle 0 | \tilde{\Phi}^\dagger_2(\phi, \eta, \chi_{1}^{I}) \,  \tilde{\Phi}_1(\phi, \eta, p\chi_{1}^{I}+q\Omega^{IJ} {\partial /  \partial \chi_1^J})  {\rm exp} ( - r \Omega_{IJ} \chi^I_1 \chi^J_1 /s ) |0\rangle.
\ee

Notice that $ \langle \gamma |\psi_j \rangle$ is just the CS path integral on $M$ with an insertion of the Wilson loop operator $W_j(C)$ along the trivial knot $C$, i.e., it is the topologically-invariant expectation value $\langle W_j(C) \rangle_{\rm CS}$ of  $W_j(C)$  in CS theory on $M$ (see also footnote~\ref{u-rep}). 

Notice also that $\langle 0 | \tilde{\Phi}^\dagger_2(\phi, \eta, \chi_{1}^{I}) \,  \tilde{\Phi}_1(\phi, \eta, p\chi_{1}^{I}+q\Omega^{IJ} {\partial /  \partial \chi_1^J})  {\rm exp} ( - r \Omega_{IJ} \chi^I_1 \chi^J_1 /s ) |0\rangle$ can be expressed as the scalar product $\langle \Phi^{(2)} | \Phi^{(1)} \rangle$ of the non-dynamically gauged RW theory on $M$ with target $X$, where $ | \Phi^{(2)} \rangle =  \tilde{\Phi}_2(\phi, \eta, \chi_{1}^{I}) | 0 \rangle$, and $| \Phi^{(1)} \rangle  = \tilde{\Phi}_1(\phi, \eta, p\chi_{1}^{I}+q\Omega^{IJ} {\partial /  \partial \chi_1^J}) \cdot {\rm exp} ( - r \Omega_{IJ} \chi^I_1 \chi^J_1 /s ) |0\rangle = \tilde \Phi'_1(\phi, \eta, \chi_{1}^{I}) |0 \rangle$. Since $\tilde \Phi'_1(\phi, \eta, \chi_{1}^{I})$ and $\tilde \Phi_2(\phi, \eta, \chi_{1}^{I})$ correspond to classes in $H_G({\cal M}^\vartheta)$, the scalar product $\langle \Phi^{(2)} | \Phi^{(1)} \rangle = \langle 0 |  \tilde \Phi^\dagger_2 \tilde \Phi'_1 | 0\rangle$ can be computed via $G$-equivariant Poinc\'are duality as an intersection number $(\tilde \Phi_2, \tilde \Phi'_1)_{{\cal M}^\vartheta}$ of $G$-equivariant cycles in ${\cal M}^\vartheta$ that are dual to $\tilde \Phi_2$ and $\tilde \Phi'_1$, respectively.

Therefore, we can actually write
\be
\boxed{Z_{X}(M^{U})=   K_0{}^j  \langle W_j(C) \rangle_{{\rm CS}(M)} \cdot (\tilde \Phi_2, \tilde \Phi'_1)_{{\cal M}^\vartheta(M, X)}}
\ee
As claimed, the partition function on $M^U$ can be calculated in terms of well-defined quantities on $M$. In fact, it can be expressed  as a product of a CS and an equivariant RW topological invariant of $M$!

\newsection{New Knot Invariants From Supersymmetric Wilson Loops}

As mentioned in subsection 4.2, a $Q$-invariant (and therefore gauge-invariant) Wilson loop operator along a knot $\mathcal{K} \subset M$ can be constructed as\footnote{Note that we have used the relation $\mu^{a}_{+}\mu_{+a}=0$ to construct the following expression.}
\begin{equation}
{W}_R(\mathcal{K})={\rm Tr}_{R}P\exp\left(\oint_{\mathcal{K}} A^a\mu_{+ a}\right),
\label{wilson1}
\end{equation}
where $M$ is an arbitrary Riemannian three-manifold, and $R$ denotes the representation $R$ of the Lie group $G$ which acts on the hyperk\"ahler target space $X$. The trace Tr is taken over $G$ whose Lie algebra $\frak g$ is generated by  the $\mu_{+ a}$'s in the representation $R$ (see footnote~\ref{u-rep}). 

\bigskip\noindent{\it The Canonical Formalism}

In the canonical formalism of section 4, where we restrict ourselves to the zero modes of the fields in the region $\Sigma \times I \subset M$ of interest, we have, in the absence of the Wilson loop operator $W_R({\cal K})$, the ``Gauss Law'' constraint $\delta L / \delta A_\tau = 0$:
\begin{equation}
\label{gauss empty}
F_{\mu\nu}^{a} = 0,
\end{equation}
where $\tau$ and $\{\mu, \nu \}$ are the coordinates on $I$ and $\Sigma$, respectively. 

If we include in our theory, multiple copies of the Wilson loop operator $W_{R_1}({\cal K}_1) W_{R_2}({\cal K}_2) \cdots$ in the representations $R_1, R_2, \dots$ of $G$, the ``Gauss Law'' constraint becomes
\begin{equation}
\label{gauss nonempty}
F_{\mu\nu}^{a}  =\epsilon_{\mu\nu}\sum_{s} \delta^2(x- P_{s})\mu_{+ (s)}^{a}.
\end{equation}
Here, the $P_s$'s are the points on $\Sigma$ that the knots ${\cal K}_1, {\cal K}_2, \dots$ intersect, and they are labeled by the representations $R_s$ via the $\mu_{+ (s)}^{a}$'s (see footnote~\ref{u-rep}). 

The physical Hilbert space ${\cal H}_{\Sigma, P_s, R_s}$ of our theory can then be obtained by quantizing the underlying symplectic phase space $\mathscr M$ determined by the $A_\mu, \eta, \chi_\tau, \chi_\mu$ fields, their respective momentum conjugate $A_\nu$, $\chi_\tau$, $\eta$, $\chi_\nu$ (computed in (\ref{cc})), the constraint (\ref{gauss nonempty}), and the conditions $D\phi = 0$, $D \chi = 0$ and $D\eta =0$. Specifically, according to the theory of geometric quantization~\cite{woodhouse}, ${\cal H}_{\Sigma, P_s, R_s}$ would correspond to the space $H^0(\mathscr L, \mathscr M)$  of holomorphic sections of a certain line bundle $\mathscr L$, where the curvature of $\mathscr L$ is given by $\sqrt {-1}$ times the symplectic two-form  of $\mathscr M$.


\bigskip\noindent{\it The Path Integral Formalism and New Knot Invariants}

Now that we have furnished, through the canonical formalism perspective, a formal description of the physical Hilbert space of the theory in the presence of multiple Wilson loop operators $W_{R_j}({\cal K}_j)$, let us compute explicitly the expectation value of such Wilson loop operators which will provide us with \emph{new} knot invariants of three-manifolds. 

To compute the expectation value $\langle W_{R}(\cal K) \rangle$ via the path integral, we will need to replace $W_{R}(\cal K)$ in (\ref{wilson1}) with its gauge-fixed version. To this end, recall that after gauge-fixing, $Q$ would be replaced by $\hat{Q}=Q+Q_{\rm FP}$. Of course,  $W_{R}(\cal K)$ in (\ref{wilson1}) is no longer invariant under the field transformations generated by $\hat{Q}$. Nevertheless, a $\hat{Q}$-invariant gauge-fixed replacement can be constructed as
\begin{equation}
\boxed{
\tilde{{W}}_R(\mathcal{K})={\rm Tr}_{R}P \exp \oint_{\mathcal{K}} \mathbb{A}:= {\rm Tr}_{R}P \exp \left( \oint_{\mathcal{K}} (A_a\mu^a_{+}+\chi^{J}\partial_{J}\mu^a_{+}c_a -\frac{1}{2}f_{abd}A^{a}c^{b}c^{d}) \right)}
\end{equation}
One can show that
\begin{equation}
\delta_{\hat{Q}}{\oint  \left(A_{a}\mu_{+}^{a}+\chi^{J}\partial_{J}\mu_{+}^{a}c_{a}-\frac{1}{2}f_{abd}A^{a}c^{b}c^{d}\right)}=\oint d\left(c_{a}\mu_{+}^{a}-\dfrac{1}{6}f_{abd}c^{a}c^{b}c^{d}\right)
=0,
\end{equation}
so
\begin{equation}
\delta_{\hat{Q}}\tilde{{W}}_R({\cal K})=0,
\end{equation}
as claimed.

To compute perturbatively the following expectation value of multiple Wilson loops 
\begin{equation}
\langle \prod_{j}\tilde{{W}}_{R_j} ({\cal K}_j) \rangle = \int D\phi DA D\eta D\chi Dc D\bar{c} \ e^{-S}\ \prod_{j}\tilde{{W}}_{R_j}({\cal K}_j),
\end{equation}
(where the Lagrange multiplier field $B$ has already been integrated out to give the gauge-fixing condition $\partial^\mu A^a_\mu = 0$), we first expand each $\tilde{{W}}_{j}$ around the flat connection $A_{0}$ and the covariantly constant map $\phi_{0}$ as
\begin{eqnarray}
\label{W-loop}
\tilde{{W}}_{R_j}({\cal K}_j) = {\rm Tr}_{R_j}P\exp\int_{\mathcal{K}_{j}} \mathbb{A}
={ \rm Tr}_{R_j}P\exp\int_{\mathcal{K}_{i}}
&&\left\{ A_{0 a}\mu^a_{+}(\phi_{0})+\tilde{A}_a\mu^a_{+}(\phi_{0})+\tilde{A}_a\partial_{J}\mu^a_{+}(\phi_{0})\varphi^{J}+\cdots \right.
 \nonumber \\ 
 &&+\chi^{J}\partial_{J}\mu^a_{+}(\phi_{0})c_a + \chi^{J}\partial_{K}\partial_{J}\mu^a_{+}(\phi_{0})\varphi^{K}c_a + \cdots  \nonumber \\ 
&& \left. -\frac{1}{2}(f_{abd}A^{a}_0c^{b}c^{d} + f_{abd}{\tilde A}^{a}c^{b}c^{d})  \right\},
\end{eqnarray}
where $\tilde A$ and $\varphi$ are fluctuations around $A_0$ and $\phi_0$, and `$\cdots$' denotes all other expansion terms around $\phi_{0}$. Notice that the above path-ordered exponential can also be expressed as
\begin{eqnarray}
\notag
\hspace{-1.5cm}{\rm Tr}_{R_j}P\exp\int_{\mathcal{K}_{j}} \mathbb{A}
 &&= {\rm Tr}_{R_j}\left(\exp\int_{\mathcal{K}_{j}}A_{0}\mu_{+}(\phi_{0})\right) \\ \notag
\hspace{-1.5cm}&&\times \left[  1 \right. \\ \notag
\hspace{-1.5cm}&&\ \ \ +{\rm Tr}_{R_j}\int_{\mathcal{K}_{j}}\left(
\tilde{A}\mu_{+}(\phi_{0})+\tilde{A}\partial_{J}\mu_{+}(\phi_{0})\varphi^{J}+
\chi^{J}\partial_{J}\mu_{+}(\phi_{0})c-\frac{1}{2}(A_{0}cc+\tilde{A}cc)+\cdots \right)  \\ \notag
\hspace{-1.5cm}&&\ \ \ + {\rm Tr}_{R_j}\int_{\mathcal{K}_{j}\times\mathcal{K}_{j}}
\left(\tilde{A}\mu_{+}(\phi_{0})+\tilde{A}\partial_{J}\mu_{+}(\phi_{0})\varphi^{J}+
\chi^{J}\partial_{J}\mu_{+}(\phi_{0})c-\frac{1}{2}( A_{0}cc+ \tilde{A}cc)+\cdots \right)^{2} \\
\hspace{-1.5cm} &&\ \ \ \left. \  +\cdots  \right ], 
\label{W expan}
\end{eqnarray}
where we have and shall henceforth omit the Lie algebra index for notational simplicity. Note that because of (\ref{moment m}), any term in the correlation function which contains ${\rm Tr}_{R_j}\left(\mu_{+}^{2}(\phi_{0})\right)$ is automatically zero.\footnote{For example, according to (\ref{moment m}), the correlation function
\begin{equation}
\langle {\rm Tr}_{R_j}(\tilde{A}\mu_{+}(\phi_{0})\tilde{A}\mu_{+}(\phi_{0})) \rangle \sim
\langle {\rm Tr}_{R_j}(\mu_{+}^{2}(\phi_{0})) \rangle =0
\end{equation}
for the classical configuration $\phi_{0}$.}

After performing the expansion, we can evaluate the correlation function of $\prod_{j}\tilde{{W}}_{k_j}({\cal K}_j)$ by the same method used to evaluate the partition function in section 3. Because of (\ref{W expan}), we can, like in (\ref{Z expan}), express the correlation function as
\begin{equation}
\langle \prod_{j}\tilde{{W}}_{K_j} ({\cal K}_j) \rangle =\sum_{A^\vartheta_0}\left(e^{-\int_{M}k_{cs}L_{cs}(A^{\vartheta}_{0})} \cdot Z_0(A^\vartheta_0)
\cdot \prod_{j}{\rm Tr}_{R_j}e^{\int_{\mathcal{K}_{j}}A_{0}^{\vartheta}\mu_{+}(\phi_{0})}  \right) 
\mathfrak{W}(M, X, G; A^{\vartheta}_{0}; k_{cs}),
\end{equation}
where the first factor in parenthesis is manifestly topologically-invariant -- $e^{-\int_{M}k_{cs}L_{cs}(A^{\vartheta}_{0})}$ is the topological factor coming from  the Chern-Simons part of the total Lagrangian evaluated at a flat connection $A_{0}^{\vartheta}$,  $Z_0(A^\vartheta_0)$ is the topological one-loop contribution given in (\ref{Z0}), and $\prod_{j}{\rm Tr}_{R_j}e^{\int_{\mathcal{K}_{j}}A_{0}^{\vartheta}\mu_{+}(\phi_{0})}$ is the product of noninteracting topological Wilson loops -- and
\begin{equation}
\mathfrak{W}(M, X, G;A_{0}^{\vartheta}; k_{cs})= \sum_{\Gamma} {\mathfrak{W}_{\Gamma}}(M, X, G; A_{0}^{\vartheta}; k^m_{cs}).
\end{equation}
Here, $\sum_{\Gamma}$ is a sum over all possible Feynman diagrams with two or more loops that (i) have the right number of fermionic zero modes to absorb those that appear in the path integral measure, and (ii) are free of the coupling constant $k$. The label $k^m_{cs}$ (where $m$ may vanish) means that $\Gamma$ carries with it a factor of $k^m_{cs}$.

Note that we have two types of Feynman diagrams here. The first type is where the vertices of $\tilde{{W}}_{k_j}({\cal K}_j)$ do not contract with the vertices of the Lagrangian $L$; let us denote this type of diagrams as $\Gamma^{\ast}$. The second type is where the vertices of $\tilde{{W}}_{k_j}({\cal K}_j)$ contract with the vertices of $L$; let us denote this type of diagrams as $\Gamma^{\diamond}$. 
In other words, we can write the total expectation value as
\begin{equation}
\langle \prod_{j}\tilde{{W}}_{K_j} ({\cal K}_j) \rangle =\langle \prod_{j}\tilde{{W}}_{R_j} ({\cal K}_j) \rangle_{\Gamma^\ast}+\langle \prod_{j}\tilde{{W}}_{R_j} ({\cal K}_j) \rangle_{\Gamma^\diamond}.
\end{equation}
Because the total expectation value $\langle \prod_{j}\tilde{{W}}_{K_j} ({\cal K}_j) \rangle$   
is topologically-invariant at the outset,  we have
\begin{equation}
\label{ind}
\frac{\delta \langle \prod_{j}\tilde{{W}}_{K_j} ({\cal K}_j) \rangle}{\delta h_{\mu\nu}} =
\frac{\delta \langle \prod_{j}\tilde{{W}}_{K_j} ({\cal K}_j) \rangle_{\Gamma^{\ast}}}{\delta h_{\mu\nu}}+
\frac{\delta \langle \prod_{j}\tilde{{W}}_{K_j} ({\cal K}_j) \rangle_{\Gamma^{\diamond}}}{\delta h_{\mu\nu}}=0,
\end{equation}
where $h_{\mu\nu}$ is the metric of $M$. 

Similar to CS and RW theory, because the propagators are not topologically-invariant, each diagram in $ \langle \prod_{j}\tilde{{W}}_{K_j} ({\cal K}_j) \rangle $ is not topologically-invariant by itself. However, the total expectation value is still topological because the variations (under a change in $h_{\mu \nu}$) of the diagrams cancel themselves out exactly. 

In our case, notice that for the diagrams $\Gamma^{\ast}$, we only have the propagator factors
\begin{equation}
\int_{M\times M} d^{3}x_{l} d^{3}y_{l} \ \Delta(x_{l},y_{l})
\end{equation}
and
\begin{equation}
\label{prop-3}
\int_{\mathcal{K}_{i}\times\mathcal{K}_{j}} dx_{l} dy_{l} \ \Delta(x_{l},y_{l}),
\end{equation}
while for the diagrams $\Gamma^{\diamond}$, we \emph{also} have the propagator factor
\begin{equation}
\label{prop-1}
\int_{M} d^{3}x_{l} \int_{\mathcal{K}_{j}} dy_{l} \ \Delta(x_{l},y_{l})
\end{equation} 
coming from the contractions between the vertices of $\tilde{{W}}_{k_j}({\cal K}_j)$ and that of the Lagrangian $L$. This means that the variations of the $\Gamma^\ast$ diagrams cannot cancel out the variations of the $\Gamma^\diamond$ diagrams. In turn, this and (\ref{ind}) imply that
\begin{equation}
\frac{\delta \langle \prod_{j}\tilde{{W}}_{K_j} ({\cal K}_j) \rangle_{\Gamma^{\ast}}}{\delta h_{\mu\nu}}=0
 \ \ \ {\rm and}\ \ \
\frac{\delta \langle \prod_{j}\tilde{{W}}_{K_j} ({\cal K}_j) \rangle_{\Gamma^{\diamond}}}{\delta h_{\mu\nu}}=0
\end{equation}
simultaneously.   In other words, both $\langle \prod_{j}\tilde{{W}}_{R_j} ({\cal K}_j) \rangle_{\Gamma^\ast}$ and $\langle \prod_{j}\tilde{{W}}_{R_j} ({\cal K}_j) \rangle_{\Gamma^\diamond}$ are \emph{independently} topologically-invariant. 

For brevity, let us henceforth focus our discussion on $ \langle \prod_{j}\tilde{{W}}_{K_j} ({\cal K}_j) \rangle_{\Gamma^{\ast}}$. For the diagrams $\Gamma^{\ast}$, we can write 
\begin{equation}
 \mathfrak{W}_{\Gamma^{\ast}}(M, X, G; A^{\vartheta}_{0}; k^m_{cs})=\int_{{\cal M}^\vartheta} \sqrt{g} \, d^{2n}\phi^{I}_{0} d^{2n}\phi^{\bar{I}}_{0} \, \,
                                  W_{\Gamma^{\ast}}(X, G;\phi_{0}, A^{\vartheta}_{0})f_{\Gamma^{\ast}}(X, G;\phi_{0}, A^{\vartheta}_{0})I^{'}_{\Gamma^{\ast}}(M;A^{\vartheta}_{0}; k^m_{cs}) \prod_{j} \Gamma^{\ast}_{{\tilde W}_{j}},
\end{equation}
where the functions $W_{\Gamma^{\ast}}$, $f_{\Gamma^{\ast}}$ and $I^{'}_{\Gamma^{\ast}}$ are similar to those in (\ref{W})--(\ref{Z-sum-2}) as they result solely from contractions among the vertices coming from $L$, and $\Gamma^{\ast}_{{\tilde W}_{j}}$ denotes the contribution of $\tilde{{W}}_{R_j}({\cal K}_j)$ to each $\mathfrak{W}_{\Gamma^{\ast}}$. 

According to the discussion leading up to (\ref{par2}), we can also write\footnote{For example, consider the contribution $\Gamma^\ast_{\tilde W_j, A \varphi}$ from the term
\begin{equation}
{\rm Tr}_{R_j}\int_{\mathcal{K}_{j}\times\mathcal{K}_{j}}
\tilde{A}\partial_{I}\mu_{+}\varphi^{I}\tilde{A}\partial_{J}\mu_{+}\varphi^{J}
\end{equation}
 in (\ref{W expan}) (assuming that the fermionic zero modes in the measure have been absorbed exactly by the vertices from $L$ which accompany this term). Performing the contraction, we get
\begin{eqnarray}
\notag
\Gamma^\ast_{\tilde W_j, A \varphi} & = &{\rm Tr}_{R_j}\int_{\mathcal{K}_{j}\times\mathcal{K}_{j}}dx^{\mu}dx^{\nu} \
\partial_{I}\mu_{+}\partial_{J}\mu_{+}
\contraction{ }{\tilde{A}_{\nu}}{\varphi^{I}}{\varphi^{J}}
\contraction[2ex]{\tilde{A}_{\mu}}{\tilde{A}_{\nu}}{\varphi^{I}}{\varphi^{J}}
\tilde{A}_{\mu}{\tilde{A}_{\nu}}{\varphi^{I}}{\varphi^{J}}
\\
&=& {\rm Tr}_{R_j}\int_{\mathcal{K}_{j}\times\mathcal{K}_{j}}dx^{\mu}dx^{\nu} \
\partial_{I}\mu_{+}\partial_{J}\mu_{+}
\Delta^{(A\varphi)I}_{\mu}\Delta^{(A\varphi)J}_{\nu} \\ \notag
&=& {\rm Tr}_{R_j}\left(\partial_{I}\mu_{+}\partial_{J}\mu_{+} \
f^{I}(X,G)f^{J}(X,G) \ \int_{\mathcal{K}_{j}\times\mathcal{K}_{j}}dx^{\mu}dx^{\nu}
\Delta^{'(A\varphi)}_{\mu}\Delta^{'(A\varphi)}_{\nu} \right) \\ \notag
& = & W_{\Gamma^{\ast}_{W_j, A \varphi}}(X,G;\phi_{0},A^{\vartheta}_{0}) \, f_{\Gamma^{\ast}_{W_j, A \varphi}}(X, G;\phi_{0}, A^{\vartheta}_{0}) \, I^{'}_{\Gamma^{\ast}_{W_j, A \varphi}}(M;A_{0}^{\vartheta}),
\end{eqnarray}
where $\mu, \nu$ run over all directions in $M$, and  the second-last equality is due to (\ref{boson propogator}). 
} 
\begin{equation}
\Gamma^{\ast}_{{W}_{j}}=W_{\Gamma^{\ast},{W}_{j}}(X,G;\phi_{0},A^{\vartheta}_{0}) \, f_{\Gamma^{\ast},{W}_{j}}(X, G;\phi_{0}, A^{\vartheta}_{0}) \, I^{'}_{\Gamma^{\ast},{W}_{j}}(M;A_{0}^{\vartheta}).
\end{equation}
Thus, we have
\be
\mathfrak{W}_{\Gamma^{\ast}}(M, X, G; A^{\vartheta}_{0}; k^m_{cs}) =  \mathscr W_{\Gamma^\ast} (X, G; A^{\vartheta}_{0}) \, {\mathscr I}^{'}_{\Gamma^\ast}(M;A^{\vartheta}_{0}; k^m_{cs}),
\ee                                 
where
\begin{equation}
\label{W-knot}
\boxed{\mathscr W_{\Gamma} (X, G; A^{\vartheta}_{0}) = \int_{\mathcal{M^\vartheta}} \, \sqrt{g} \, d^{2n}\phi^{I}_{0} \, d^{2n}\phi^{\bar{I}}_{0}
                                 \,\, (W_{\Gamma^\ast}f_{\Gamma^\ast} \prod_j W_{\Gamma^{\ast},{W}_{j}} f_{\Gamma^{\ast},{W}_{j}}) (X, G;\phi_{0}, A^{\vartheta}_{0})}
\end{equation}
can be regarded as a weight factor which combines the structural information of the hyperk\"ahler manifold $X$ and the Lie algebra  $\mathfrak{g}$ of the gauge group $G$, and
\be
\boxed{{\mathscr I}^{'}_{\Gamma^\ast}(M;A^{\vartheta}_{0}; k^m_{cs}) = (I^{'}_{\Gamma^{\ast}} \prod_j I^{'}_{\Gamma^{\ast},{W}_{j}})(M;A_{0}^{\vartheta}; k^m_{cs})}
\ee
Therefore, 
\begin{equation}
 \label{knots-nonint}
 \langle \prod_{j}\tilde{{W}}_{R_j} ({\cal K}_j) \rangle_{\Gamma^\ast}= \sum_{A^\vartheta_0} \sum_{\Gamma^\ast} \left(e^{-\int_{M}k_{cs}L_{cs}(A^{\vartheta}_{0})} \cdot Z_0
\cdot \prod_{j}{\rm Tr}_{R_j}e^{\int_{\mathcal{K}_{j}}A_{0}^{\vartheta}\mu_{+}(\phi_{0})}  \right)    \mathscr W_{\Gamma^\ast} (X, G; A^{\vartheta}_{0}) \, {\mathscr I}^{'}_{\Gamma^\ast}(M;A^{\vartheta}_{0}; k^m_{cs}),
 \end{equation}
where 
\begin{equation}
 \label{knot-inv-final}
 \boxed{\langle \prod_{j}\tilde{{W}}_{R_j} ({\cal K}_j) \rangle_{\Gamma^\ast, A^\vartheta_0}=  \sum_{\Gamma^\ast}  \,   \mathscr W_{\Gamma^\ast} (X, G; A^{\vartheta}_{0}) \, {\mathscr I}^{'}_{\Gamma^\ast}(M;A^{\vartheta}_{0}; k^m_{cs})}
 \end{equation}
is a \emph{new} knot invariant of three-manifolds which depends on both $G$ and $X$, that also defines a \emph{new} knot weight system whose weights $ \mathscr W_{\Gamma^\ast} (X, G; A^{\vartheta}_{0})$ are characterized by both Lie algebra structure    \emph{and} hyperk\"ahler geometry.

\end{document}